\documentclass[preprint,
superscriptaddress,
amsmath,
amssymb,
aps
]{revtex4-1}

\usepackage{graphicx}
\usepackage{dcolumn}
\usepackage{bm}
\usepackage{xcolor}

\begin{document}

\title{Electric Field Induced Associations in the Double Layer of Salt-in-Ionic-Liquid Electrolytes}

\author{Daniel M. Markiewitz}
\affiliation{Department of Chemical Engineering, Massachusetts Institute of Technology, Cambridge, Massachusetts  02139, USA}

\author{Zachary A. H. Goodwin}
\affiliation{John A. Paulson School of Engineering and Applied Sciences, Harvard University, Cambridge, Massachusetts 02138, United States}
\affiliation{Departments of Materials, Imperial College London, South Kensington Campus, London SW7 2AZ, UK}

\author{Michael McEldrew}
\affiliation{Department of Chemical Engineering, Massachusetts Institute of Technology, Cambridge, Massachusetts  02139, USA}

\author{J. Pedro de Souza}
\affiliation{Department of Chemical Engineering, Massachusetts Institute of Technology, Cambridge, Massachusetts  02139, USA}
\affiliation{Omenn-Darling
Bioengineering Institute, Princeton University, Princeton, New Jersey 08544, USA}

\author{Xuhui Zhang}
\affiliation{Department of Civil and Environmental Engineering, University of Illinois at Urbana–Champaign, Urbana, IL, 61801 USA}

\author{Rosa M. Espinosa-Marzal}
\affiliation{Department of Civil and Environmental Engineering, University of Illinois at Urbana–Champaign, Urbana, IL, 61801 USA}
\affiliation{Department of Materials Science and Engineering, University of Illinois at Urbana–Champaign, Urbana, IL, 61801 USA}

\author{Martin Z. Bazant}
\email{e-mail: bazant@mit.edu}
\affiliation{Department of Chemical Engineering, Massachusetts Institute of Technology, Cambridge, Massachusetts  02139, USA}
\affiliation{Department of Mathematics, Massachusetts Institute of Technology, Cambridge, Massachusetts 02139, USA}

\date{\today}

\begin{abstract}

Ionic liquids (ILs) are an extremely exciting class of electrolytes for energy storage applications because of their unique combination of properties. Upon dissolving alkali metal salts, such as Li or Na based salts, with the same anion as the IL, an intrinsically asymmetric electrolyte can be created for use in batteries, known as a salt-in-ionic liquid (SiIL). These SiILs have been well studied in the bulk, where negative transference numbers of the alkali metal cation have been observed from the formation of small, negatively charged clusters. The properties of these SiILs at electrified interfaces, however, have received little to no attention. Here, we develop a theory for the electrical double layer (EDL) of SiILs where we consistently account for the thermoreversible association of ions into Cayley tree aggregates. The theory predicts that the IL cations first populate the EDL at negative voltages, as they are not strongly bound to the anions. However, at large negative voltages, which are strong enough to break the alkali metal cation-anion associations, these IL cations are exchanged for the alkali metal cation because of their higher charge density. At positive voltages, we find that the SiIL actually becomes \textit{more aggregated while screening the electrode charge} from the formation of large, negatively charged aggregates. Therefore, in contrast to conventional intuition of associations in the EDL, SiILs appear to become more associated in certain electric fields. We present these theoretical predictions to be verified by molecular dynamics simulations and experimental measurements.

\end{abstract}

\maketitle

\section{Introduction}

Alkali metal salt doped ionic liquids, referred to as salt-in-ionic liquids (SiILs) here, have attracted attention as an electrolyte for applications in batteries/energy storage devices~\cite{watanabe2017application,lewandowski2009ionic,balducci2004ionic,dokko2013solvate,xu2014electrolytes,Chagas2019,Zhou2011,monti2014battery,armand2009ionic,matsumoto2019advances,mendes2018ionic,zhao2019effect,yamamoto2023amide,guo2024alkali,basile2018ionic,macfarlane2014energy,macfarlane2016ionic}, owing to the combination of the active Li/Na cations coupled with the highly desirable properties of ionic liquids (ILs). ILs have extremely low vapour pressures, are effectively universal solvents, and are non-flammable making them a much safer alternative to carbonate based solvents typically used in conventional battery electrolytes~\cite{Welton1999,Hermann2008,Welton2011,Fedorov2014}. Unlike conventional battery electrolytes, SiILs are solely composed of ions, where the anion of the alkali metal salt and IL are often the same, e.g., Li[TFSI] dissolved in [EMIM][TFSI] resulting in there being more anions than either type of the cation. Therefore, these electrolytes are \textit{inherently asymmetric}, and moreover, strongly correlated because of the super-concentration of salts~\cite{McEldrewsalt2021}. 

One area where the asymmetry and correlation manifests is in the transport properties of the SiILs~\cite{McEldrewsalt2021}. In electrophoretic NMR experiments~\cite{qiao2018supramolecular,gouverneur2015direct,gouverneur2018negative,Marc2021,Ackermann2021} and MD simulations~\cite{molinari2019transport,molinari2019general,chen2016elucidation,Kubisiak2020,lourenco2021theoretical,molinari2020chelation,Haskins2014} of SiILs, it has been revealed that, at low mole fractions of the alkali metal salt, the alkali metal cation has a \textit{negative} transference number, indicating that it moves in the opposite direction of the electric field. This property has been explained by the alkali metal cation being strongly solvated by anions. Owing to the larger number of anions, these species can form small, but high negatively charged clusters that are transported in a vehicular way~\cite{molinari2019transport,molinari2019general,McEldrewsalt2021}. 

These strong interactions are also evident at higher mole fractions. With increasing mole fraction, the sizes of the clusters increase from only containing 1 cation to containing several cations in an extended cluster connected by anions, as found to occur in MD simulations~\cite{molinari2019transport,molinari2019general,ren2023molecular}. These clusters continue to increase in size with the mole fraction and, eventually, a percolating ionic network of alkali metal cations and anions occurs~\cite{molinari2019transport,molinari2019general}. A simple, analytical theory for the ionic clusters of SiILs was developed by McEldrew \textit{et al.}~\cite{McEldrewsalt2021}, based on the analogy between polymers and concentrated electrolytes~\cite{Goodwin2023}, which was able to rationalise the changes to the cluster distribution with mole fraction~\cite{McEldrewsalt2021}. Drawing from the famous theories of Flory, Stockmayer and Tanaka~\cite{mceldrew2020theory}, it was predicted that the percolating ionic network of ions should behave like a gel~\cite{mceldrew2020theory,mceldrew2020corr,mceldrew2021ion}, and in fact, experimental indications of this gel have been found in water-doped SiILs~\cite{reber2021anion}. Therefore, it is clear that ionic aggregation in SiILs is an extremely important phenomena.

One aspect of SiILs that is not well understood at all, however, is the electrical double layer (EDL), i.e. how the ions arrange themselves at an electrified interface. To our knowledge, Haskins \textit{et al.}~\cite{Haskins2016} is one of the only papers investigating this topic in detail, presumably because of the complexity of the electrolyte. It found that the Li$^+$ induces disorder in the EDL and do not prefer to reside right at the electrode interface, opting instead to reside within the anion-rich layers. Recently, Goodwin \textit{et al.}~\cite{Goodwin2022EDL,Goodwin2022Kornyshev} extended the theories of McEldrew \textit{et al.}~\cite{mceldrew2020theory,mceldrew2020corr} to make predictions in the EDL finding that the electric field induces cracking of clusters~\cite{Goodwin2022EDL,Goodwin2022Kornyshev}. However, this work has only been applied to simple symmetric ILs and further development is necessary to extend the theories of McEldrew and Goodwin \textit{et al.} to SiILs.

\begin{figure}[h!]
 \centering
 \includegraphics[width= 0.7\textwidth]{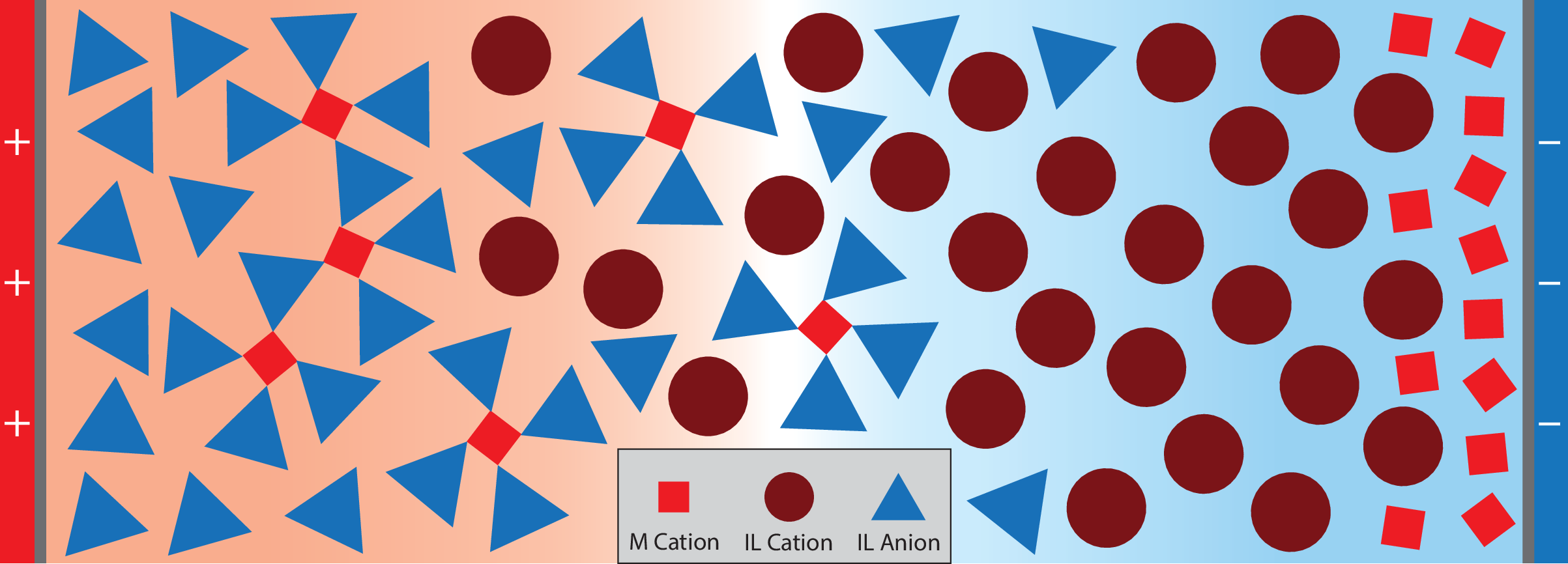}
 \caption{Schematic of the modulation of aggregations occurring in the EDL of SiILs near a positively (left) and negatively (right) charged electrodes. Here the metal cations can form up to 4 associations and the IL anions up to 3 associations. Ion associations are shown by touching vertices.}
 \label{fig:vis_edl}
\end{figure}

In this paper, we develop a theory for the EDL of SiILs based on the work of McEldrew and Goodwin and co-workers. Conventional theories for the EDL are not able to correctly capture the associations which occur in the EDL~\cite{goodwin2017mean,Chen2017,Goodwin2017,Yufan2020} nor the extended ionic aggregates which occur in SiILs~\cite{molinari2019general,McEldrewsalt2021}. Here we present a theory for the EDL of SiILs which accounts for ionic associations in a consistent way and allows us to investigate these associations in the EDL of a strongly correlated and intrinsically asymmetric electrolyte. We find two main observations from this theory. Firstly, a cation exchange occurs at negative potentials, where the IL cation is found to be dominant at small negative potentials but is then replaced by the smaller alkali metal cation at increasingly negative potentials. This exchange occurs because of the competition between different sizes of species and the associations between the alkali metal cations and anions. Secondly, in positive potentials, we find there is some intermediate range of potentials where the SiIL \textit{becomes more associated than in the bulk}. These results are schematically shown in Fig.~\ref{fig:vis_edl}.

\section{Theory}
We consider our system to be an incompressible lattice-gas model ~\cite{mceldrew2020theory}, which is composed of alkali metal cations, denoted by $+$, and IL cations and anions, denoted by, respectively, $\oplus$ and $-$. In the bulk, the total volume fractions, $\phi_j$, of each species are known, where $j$ is $+$, $-$ or $\oplus$. The lattice site is assumed to be the size of the alkali metal cation, $v_+$, with the volumes occupied by the other species being determined by $v_+ \xi_i$, where $\xi_i = v_i/v_+$ with $i$ being $-$ or $\oplus$ here. The dimensionless concentrations of each species are given by $c_j = \phi_j/\xi_j$, with $\xi_+ = 1$ for the alkali metal cations. 

Similar to previous work, it is assumed that the anions and metal cations may form associations; but, the IL cations only interact with the open association sites of anions through a regular solution term~\cite{McEldrewsalt2021}. The metal cations can form a maximum of $f_{+}$ associations and the anions a maximum of $f_{-}$ associations, referred to as the functionality of each species. \textcolor{black}{The number of associations is typically bounded by the integer number of coordination of an associating species within the first solvation shell of each species. Typically these associations are determined via a cutoff distance between the key moieties in the molecules, such as the oxygens' in a TFSI anion and a lithium ion~\cite{mceldrew2021ion}. Additionally, they can be determined by studying the spatial distribution functions for the associating species to identify how many ``hot-spot" regions exist for the species, which corresponds to the functionality of that species~\cite{mceldrew2020corr}. Alternatively, kinetic criteria~\cite{feng2019free} or machine learning methods~\cite{jones2021bayesian} can be used to define associations between ions in electrolytes.} As these functionalities are larger than 1, a polydisperse cluster distribution can form clusters of rank $lm$, where $l$ is the number of cations and $m$ is the number of anions in \textcolor{black}{a cluster with rank} $lm$, with dimensionless concentration $c_{lm}$, but we assume that only Cayley-tree-like clusters can form, i.e. clusters with no loops~\cite{mceldrew2020theory}. This Cayley tree assumption of the clusters is required for our approach to remain analytical and physically transparent~\cite{mceldrew2020theory}, but the approximation is known to break down sometimes~\cite{McEldrewsalt2021,mceldrew2021ion}. 

For functionalities larger than 2, a percolating ionic network can emerge~\cite{mceldrew2020theory}. This is referred to as the gel here. In the gel regime, we employ Flory's convention to determine the volume fractions of each species in the sol ($\phi_{+/-}^{sol}$, which contains the clusters and free species) and gel phase ($\phi_{+/-}^{gel}$, which only contains the percolating network of ions), where $\phi_{+/-} = \phi_{+/-}^{sol} + \phi_{+/-}^{gel}$. The total dimensionless concentration of each species is given by a sum over all possible clusters, $c_+ = \sum_{lm} l c_{lm} + c_+^{gel}$, $c_- = \sum_{lm} m c_{lm} + c_-^{gel}$.

The free energy functional ($\mathcal{F}$) is proposed to take the following form

\begin{align}
    \label{HEnergy} 
    v_+\mathcal{F} =& \int_V  \,d\textbf{r} -\frac{v_+\epsilon_0\epsilon_r}{2}\big(\nabla\Phi\big)^2 + v_+\rho_e \Phi + k_BTc_{\oplus} \ln \phi_{\oplus} +\sum_{lm} \left(k_BTc_{lm}\ln\phi_{lm} + c_{lm}\Delta_{lm}\right) \nonumber \\
    &+ \int_V  \,d\textbf{r}\chi\phi_{\oplus}\sum_{lm}\left(f_- m -m-l+1\right)c_{lm} + \Delta^{gel}_+c_+^{gel} + \Delta^{gel}_-c_-^{gel}  \nonumber \\
    &+ \int_V  \,d\textbf{r} k_BT\Lambda \left(1-\xi_{\oplus}c_{\oplus}-\sum_{lm}(l+\xi_-m)c_{lm} \right) 
\end{align}

\noindent Here the electrostatic potential, $\Phi(\textbf{r})$, charge density, $\rho_e(\textbf{r})$, and volume fractions/dimen\-sionless concentrations, $\phi(\textbf{r})/c(\textbf{r})$, all vary in space from the interface and we integrate over the entire electrolyte region. The first two terms represent the electrostatic energy, where $\epsilon_0$ and $\epsilon_r$ are, respectively, the permittivity of free space and the relative dielectric constant, $\Phi$ is the electrostatic potential and $\rho_e$ is the charge density, given by $\rho_e = \frac{e}{v_+}(c_{\oplus}+c_+-c_-)$, with $e$ denoting the elementary charge. The third term is the ideal entropy of the IL cations, and the fourth term is the ideal entropy from clusters of rank $lm$. The fifth term is the free energy of forming clusters, where $\Delta_{lm}$ is the free energy of forming each cluster of rank $lm$, with more details following later. The sixth term is the regular solution interaction, with strength $\chi$, between the IL cations and the open association sites, where we find the following identity to be useful

\begin{equation}
    \label{R1}
    \sum_{lm}(f_-m-m-l+1)c_{lm} = f_-c_{-}(1 - p_{-+}).
\end{equation}

\noindent Note that this expression only holds in the pre-gel regime; where the left hand side assumes free anions or anions in clusters can interact with IL cations but the right hand side assumes all anions can interact with IL cations, even those in the gel. The seventh and eighth terms come from the free energy of species associating to the gel, $\Delta_j^{gel}$, which is a function of $\phi_{\pm}$ for thermodynamic consistency. Finally, the last term is a Lagrange multiplier to enforce the incompressibility more explicitly than previously reported~\cite{Goodwin2022EDL,Goodwin2022Kornyshev}. Note this term introduces another unknown, $\Lambda$, which needs to be found~\cite{gongadze2013spatial}. 

We consider the free energy of forming a cluster of rank $lm$ to have two contributions

\begin{equation}
    \Delta_{lm} = \Delta^{comb}_{lm} + \Delta^{bind}_{lm},
\end{equation}

\noindent where the first term is the combinatorial contribution from the number of ways of arranging the ions in each cluster and the second term is the binding free energy of each cluster. In the context of polymers, Stockmayer~\cite{stockmayer1943theory} solved the combinatorial entropy for Cayley tree associations

\begin{equation}
    \Delta_{lm}^{comb} = k_BT\ln\{f_+^lf_-^m W_{lm}\},
\end{equation}

\noindent where

\begin{equation}
    W_{lm}=\frac{(f_{+}l-l)!(f_{-}m-m)!}{l!m!(f_{+}l-l-m+1)!(f_{-}m-m-l+1)!}.
\end{equation}

\noindent Moreover, the binding free energy is simply given by

\begin{equation}
    \Delta_{lm}^{bind} = (l+m-1)\Delta f_{\pm},
\end{equation}

\noindent owing to the assumption of Cayley tree clusters.

We can calculate the chemical potential of the bare ions in the bulk and the EDL, where an overbar will indicate that the variable is the EDL version and $\Phi$ is non-zero. For the IL cations, the electrochemical potential, up to arbitrary constants, is given by

\begin{equation}
    \label{c1_pot}
    \beta \bar{\mu}_{\oplus} = \beta e \Phi +\ln\bar{\phi}_{\oplus} +\beta \chi \xi_{\oplus} f_-\bar{c}_-\left(1 - \bar{p}_{-+}\right) - \xi_{\oplus}\Lambda, 
\end{equation}

\noindent and for the clusters the chemical potential, up to arbitrary constants, is given by

\begin{equation}
    \label{clm_pot}
    \beta \bar{\mu}_{lm} = (l-m)\beta e \Phi +\ln\bar{\phi}_{lm} + \Delta_{lm} + \beta \chi \bar{\phi}_{\oplus} (f_-m-m-l+1) + (l + \xi_-m)(d' - \Lambda),
\end{equation}

\noindent where $d' = \bar{c}_+\partial \Delta_+^{gel} + \bar{c}_-\partial \Delta_-^{gel}$, with the derivative being with respect to $\bar{\phi}_{\pm} = \bar{\phi}_+ + \bar{\phi}_-$.

In the bulk by asserting the clusters are in equilibrium with the bare species

\begin{equation}
    \label{Cluster_equil_eq}
    l\mu_{10} + m\mu_{01} = \mu_{lm},
\end{equation}

\noindent we can predict the cluster distribution from the bare species

\begin{equation}
    c_{lm}=\frac{W_{lm}}{\lambda}  \left(\lambda f_+\phi_{10}\right)^l \left(\lambda f_- \phi_{01}/\xi_-\right)^m,
\label{eq:clust}
\end{equation}

\noindent where $\lambda$ is the ionic association constant given by
\begin{equation}
    \label{eq:L}
    \lambda=\exp \left\{\beta\left(-\Delta f_{\pm} + \chi \phi_{\oplus}\right)\right\} = \lambda_0\exp \left\{\beta\chi \phi_{\oplus}\right\}.
\end{equation}

By establishing the chemical equilibrium between free species and clusters \textit{within} the EDL
\begin{equation}
l\bar{\mu}_{10} + m\bar{\mu}_{01} = \bar{\mu}_{lm},
\end{equation}

\noindent we obtain an analogous solution for the cluster distribution given the volume fractions of the bare species

\begin{equation}
    \bar{c}_{lm}=\frac{W_{lm}}{\bar{\lambda}}  \left(\bar{\lambda} f_+\bar{\phi}_{10}\right)^l \left(\bar{\lambda}f_- \bar{\phi}_{01}/\xi_-\right)^m,
\end{equation}

\noindent where

\begin{equation}
    \label{eq:barL}
    \bar{\lambda} = \lambda_0\exp \left\{\beta\chi \bar{\phi}_{\oplus}\right\}.
\end{equation}

Following the work of Goodwin \textit{et al.}~\cite{Goodwin2022EDL}, we connect the bulk and EDL cluster distributions to the Poisson equation through closure relations. These closure relations are based on the pre-gel regime and naively extended to the post-gel regime (this assumes the additional chemical potential contribution is not significant, with further research being required to quantify its contribution). Equating the chemical potential of the free ions in bulk and in the EDL allows us to write down 3 additional equations. First, for the alkali metal cations

\begin{equation}
\bar{\phi}_{10} = \phi_{10}\exp(-e\beta\Phi + \Lambda),
\label{eq:closure_1}
\end{equation}

\noindent there is only contributions from the electrostatic potential and excluded volume effects. Second, for the anions

\begin{equation}
\bar{\phi}_{01} = \phi_{01}\exp(e\beta\Phi- \beta\chi f_- (\bar{\phi}_{\oplus}-\phi_{\oplus}) + \xi_-\Lambda),
\label{eq:closure_2}
\end{equation}

\noindent there is an additional contribution from the regular solution interaction with IL cations, which causes an enrichment of anions where the IL cations are. Finally, for the IL cations

\begin{equation}
\bar{\phi}_{\oplus} = \phi_{\oplus}\exp\left(-e\beta\Phi + \beta\chi f_- \xi_{\oplus}\left\{c_{-}(1 - p_{-+})-\bar{c}_{-}(1 - \bar{p}_{-+})\right\} + \xi_{\oplus}\Lambda\right),
\label{eq:closure_3}
\end{equation}

\noindent where there is also an additional term, which pulls the IL cations to where there are open association sites of the anions. Note in previous versions~\cite{Goodwin2022EDL,Goodwin2022Kornyshev}, an additional parameter, $\alpha$~\cite{goodwin2017mean}, was used in the closure relations which accounts for additional short-ranged correlations between ions. It is known these simple mean-field theories underestimate the correlations between ions, which causes the screening lengths to be too small, and this parameter is a way of correcting this deficiency. This $\alpha$ parameter ~\cite{goodwin2017mean} could also be introduced here, but for simplicity, we won't show results with this parameter.

Lastly to allow us to solve these Boltzmann closure relationships, we need to introduce the idea of association probabilities, conservation of associations and the law of mass action on associations~\cite{mceldrew2020theory}. As $\phi_{10}$ and $\phi_{01}$ are, in principle, experimentally inaccessible. Instead, it is natural to express the cluster distribution in terms of the overall volume fractions of each species, $\phi_i$, which is an experimentally and computationally controllable parameter through the mole fraction of the alkali metal salt~\cite{mceldrew2020theory}. This connection is established by introducing ion association probabilities, $p_{ij}$, which is the probability that an association site of species $i$ is bound to species $j$, where $i$ and $j$ are either the alkali cation ($+$) or the IL anion ($-$)~\cite{McEldrewsalt2021}. Therefore, the volume fraction of free alkali cations can be written as $\phi_{10} = \phi_{+}(1 - p_{+-})^{f_{+}}$ and free anions as $\phi_{01} = \phi_-(1 - p_{-+})^{f_-}$~\cite{McEldrewsalt2021}. 

The association probabilities can be determined through the conservation of associations and a mass action law between open and occupied association sites~\cite{mceldrew2020theory,McEldrewsalt2021}. The conservation of associations is given by

\begin{align}
p_{+-}\psi_{+}=p_{-+}\psi_-,
\label{eq:cons}
\end{align} 

\noindent where $\psi_{+}=f_{+}\phi_{+}$ and $\psi_{-}=f_{-}\phi_{-}/\xi_-$ are the number of alkali cation  and anions association sites per lattice site, respectively~\cite{mceldrew2020theory,McEldrewsalt2021}. The mass action law between open and occupied association sites is

\begin{align}
\lambda \zeta = \dfrac{p_{+-}p_{-+}}{(1-p_{+-})(1-p_{-+})},
\label{eq:MAL}
\end{align}

\noindent where $\zeta=\psi_- p_{-+}=\psi_{+}p_{+-}$ is dimensionless concentration of associations (per lattice site)~\cite{mceldrew2020theory,McEldrewsalt2021}. The definition of $\lambda$ as the ionic \emph{association constant} becomes clear from its appearance in the association mass action law. It sets the equilibrium for association sites to be occupied or open. Equations \eqref{eq:cons} and \eqref{eq:MAL} permit an explicit solution for the probabilities $p_{+-}$ and $p_{-+}$ in terms of overall species volume fractions

\begin{align}
\psi_- p_{-+}=\psi_{+}p_{+-}=\frac{1 + \lambda (\psi_- + \psi_{+}) - \sqrt{\left[1 + \lambda (\psi_- + \psi_{+})\right]^2-4 \lambda^2 \psi_- \psi_{+}} }{2 \lambda}.
\label{eq:p}
\end{align}

\noindent Note, analogous expression for the association probabilities occur in the EDL but where EDL quantities are utilised. 

As shown in the works of McEldrew \textit{et al.}~\cite{mceldrew2021ion,McEldrewsalt2021}, the association constants can be found from fitting Eq.~\eqref{eq:MAL} using association probabilities obtained from MD simulations. For the specific case of SiILs, there are two contributions to the association constant, $\lambda_0$ and $\chi$, which can be extracted if various salt compositions are studied~\cite{McEldrewsalt2021}. As the association constant in the EDL only depends on these bulk parameters, and variables in the EDL, we do not need to further determine $\bar{\lambda}$. In Ref.~\citenum{McEldrewsalt2021}, it was found that $\lambda_0 \approx 50$ and $\chi \approx -2$ k$_B$T, which shall mainly be used in this study. Therefore, our theory has no free fitting parameters in terms of the associations in the SiILs. \textcolor{black}{This feature is achieved as all the parameters needed to predict the EDL structure can be determined from the bulk SiIL. Explicitly one can extract $\lambda_0$, $\chi$, and integer bounds on $f_+$ and $f_-$ from MD simulations of the SiIL system at various salt compositions and the volume ratios are backed out by molecular data of the species composing the SiIL. Alternatively, one can use integral equations and Wertheim's formalism to determine the association constants without fitting association constants from MD simulations~\cite{wertheim1984fluids1,wertheim1984fluids2,wertheim1986fluids1,wertheim1986fluids2,laria1990cluster,blum1995general,simonin1999ionic,sciortino2007self}. However, this is only accessible for certain cases, such as charged hard spheres with known Mayer f-functions for interacting species, and not for specific chemistries~\cite{Goodwin2023}.} 

Taking the functional derivative of the free energy with respect to the electrostatic potential, we arrive at the Poisson equation,

\begin{equation}
    \label{PB}
    \epsilon_0\epsilon_r\nabla^2\Phi = -\rho_e = -\frac{e}{v_+}(\bar{c}_{\oplus}+\bar{c}_+-\bar{c}_-).
\end{equation}

\noindent From this equation, we can define the inverse Debye length, $\kappa = \sqrt{e^2\beta(c_{\oplus}+c_++c_-)/v_+\epsilon_0\epsilon_r}$, which will be used to to convert distances from the electrode to dimensionless values. The Poisson-Boltzmann (PB) equation follows from a mean-field approximation of Coulomb interactions, which is technically only valid in the dilute limit of point-like ions in a uniform dielectric continuum ~\cite{guldbrand1984electrical,kjellander1986interaction,grochowski2008continuum}. Many modified PB equations are available which involve corrections for finite ion sizes, Coulomb correlations, and non-local dielectric response~\cite{netz2001electrostatistics,bazant2009a,Bazant2011,Pedro2020,pedro2022polar,avni2020charge,adar2019screening,gongadze2013spatial}. In our simple model, such corrections are captured indirectly through their short-range associations, which is to promote the formation of ionic clusters with ``spin glass” ordering favoring oppositely charged neighbors~\cite{levy2019spin}.

To solve the system of equations and the Poisson equation, we use the following procedure. First, we calculate bulk properties from Eq.~\eqref{eq:p} using the association constant. Next we solve our closure relations, Eqs.~\eqref{eq:closure_1}-\eqref{eq:closure_3}, with Eq.~\eqref{eq:p} and the incompressibility condition in the EDL ($1 = \bar{\phi}_+ + \bar{\phi}_- + \bar{\phi}_{\oplus}$), as we need 6 equations to solve for our 6 unknowns, to obtain the relationships between $\bar{\phi}_+,\bar{\phi}_-,\bar{\phi}_{\oplus}$ and $\Phi$, on a regular grid of $\Phi$. We then use this mapping between $\Phi$ and $\bar{\phi}$'s to numerically solve the Poisson equation. Note that this procedure is similar to that of Ref.~\citenum{Goodwin2022EDL}, but where we now also solve the Lagrange multiplier, $\Lambda$, to account for the different sizes of each species. 

An informative quantity to understanding the EDL can be the differential capacitance, C, also known as the double layer capacitance,

\begin{equation}
    \label{DC}
    C = \frac{d q_s}{d \Phi_s}.
\end{equation}

\noindent Here, q$_s$ is the surface charge at the interface and $\Phi_s$ is the electrostatic potential at the charged interface, this is equivalent to the potential drop across the EDL. Using our previous procedure to solve the Poisson equation over range of q$_s$ we can numerically solve for how the q$_s$ is a function of $\Phi_s$. Following this, we can numerically take the derivative of q$_s$ with respect to $\Phi_s$ to determine the differential capacitance. For additional details on setting up the system of equations and reproducing the calculations see the Supplemental Material.

\section{Results}

Here we describe the predictions of the model in detail. As the theory heavily relies on previous work for SiILs in the bulk, we suggest the reader is first acquainted with Ref.~\citenum{McEldrewsalt2021}. Moreover, reading Ref.~\citenum{Goodwin2022EDL} might help the reader put into context the predictions for the EDL. In all of the results shown in the main text, we choose the following parameters. A mole fraction of alkali metal salt of $\textcolor{black}{x_s} = 0.01$ is chosen, which is quite small, and was motivated by ensuring that no gelation occurs for all potentials for the rest of the parameters used. For the functionalities, we choose $f_+=5$ and $f_-=3$, inspired by previous work~\cite{McEldrewsalt2021}. We have chosen $\chi = -2$ k$_B$T and $\lambda_0=50$, which correspond to values previously used for SiILs~\cite{McEldrewsalt2021}. Finally, we choose  $\xi_+$=1, $\xi_-$=7, $\xi_{\oplus}$=7. Some predictions of the model highly depend on these parameters. We have included calculations in the Supplemental Material for how the results depend on these parameters. 

\begin{figure}[h!]
     \centering
     \includegraphics[width= 1\textwidth]{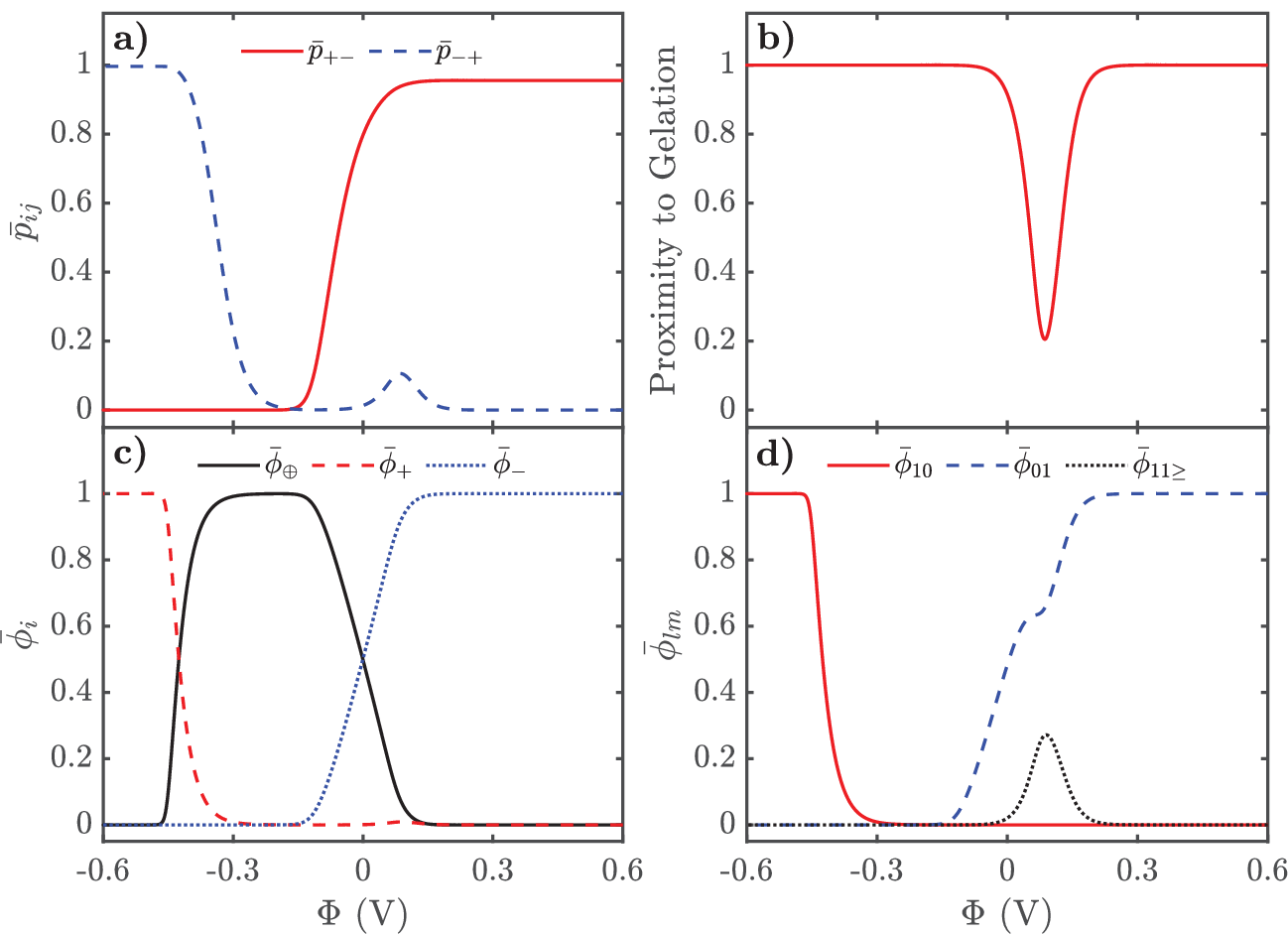}
     \caption{Properties of the EDL of SiILs as a function of applied electrostatic potential. \textcolor{black}{a)} Association probabilities\textcolor{black}{. b)} Proximity to gelation, 1-$\bar{p}_{+-}\bar{p}_{-+}$(f$_+$-1)(f$_-$-1)\textcolor{black}{. c)} Total volume fractions of each species\textcolor{black}{. d)} Volume fraction of free cations, anions and clusters. Here we use $x\textcolor{black}{_s} = 0.01$, $f_+=5$, $f_-=3$, $\xi_+=1$, $\xi_-=7$, $\xi_{\oplus}=7$, $\chi = -2$ k$_B$T, and $\lambda_0=50$.}
     \label{fig:comp_001}
\end{figure}

In Fig.~\ref{fig:comp_001}, we display the solutions to our equations as a function of electrostatic potential. \textcolor{black}{The displayed plot} does not correspond to an actual physical situation\textcolor{black}{;} but it allows us to first understand the response of SiILs to applied potentials before describing how all of these changes are distributed in space as a function of distance from the electrified interface. First, we focus on the changes in the total volume fractions of each species, $i = \oplus, +, -$, as a function of potential as seen in Fig.~\ref{fig:comp_001}\textcolor{black}{.c)}. For positive applied potentials, there is a large increase in $\bar{\phi}_-$, which plateaus at 1 at 0.2$~$V, while the IL cations, $\bar{\phi}_{\oplus}$, are monotonically expelled from the EDL. Interestingly, there is an increase in the volume fraction of alkali metal cations, $\bar{\phi}_+$, at modest applied positive potentials, before the volume fraction decreases to 0 at large applied positive voltages. For negative potentials, the anions are monotonically expelled from the EDL but the response of the cations is more complex. At small to moderate voltages, the IL cation dominates the EDL, and saturates at $\bar{\phi}_{\oplus} \approx 1$ at a potential of -0.25$~$V. At potentials more negative than -0.3$~$V, a cation exchange occurs, where the IL cations are replaced by metal cations, $\bar{\phi}_{\oplus} \approx 0$ and $\bar{\phi}_{+} \approx 1$. 

A cation exchange occurs at negative voltages because of the competition between the different sizes of the cations, and \textcolor{black}{its position is dependent on} the strong associations between the metal cations and the anions. At small potentials, the IL cations first populate the EDL because they are not strongly interacting with the anions, which are pushed out of the EDL and take the alkali metal cations with them through the associations. At large negative voltages, it would be expected that the smallest cation populates the EDL, as this maximises the charge density in the EDL and reduces its length. Therefore, it is expected that the smaller alkali metal cation saturates the EDL at large negative voltages instead of the larger IL cation. If both cations are the same size, this cation exchange does not occur as the IL cation remains in the EDL. There is a critical voltage at which this cation exchange occurs, which depends not only on the sizes of the cations but also the association constant between the alkali metal cation and the anion. For larger association constants, the alkali metal cation is more strongly bound to the anions, which means it requires larger fields to break these associations to obtain the cation exchange. In the Supplemental Material we show how the cation exchange depends on these parameters in more detail.

Next we describe the association probabilities as a function of applied voltage, as seen in Fig.~\ref{fig:comp_001}\textcolor{black}{.a)}. At large negative potentials, $\bar{p}_{+-} = 0$ and $\bar{p}_{-+} = 1$, which is expected if there is $\bar{\phi}_+ = 1$ and $\bar{\phi}_- = 0$. This can be understood by considering trace amounts of IL anions in alkali metal cations. In this case, it would expected the alkali metal picked at random to rarely be associated to an IL anion; but, any IL anion picked at random would be associated with alkali metal cations. Similarly, for large positive potentials $\bar{p}_{-+} = 0$ and $\bar{p}_{+-} = 1$ because of $\bar{\phi}_- = 1$ and $\bar{\phi}_+ = 0$. These limits are also discussed in detail in Ref.~\citenum{Goodwin2022EDL}. At a voltage of $\sim0.1~$V, however, there is also a peak in $\bar{p}_{-+}$, which occurs because there is an increase in the volume fraction of alkali metal cations. These cations are brought to the EDL from the strong associations with the anions, and therefore, results in \textit{electric field induced associations in the EDL}. This can be further seen in Fig.~\ref{fig:comp_001}\textcolor{black}{.b)}, which displays the proximity to gelation, 1-$\bar{p}_{+-}\bar{p}_{-+}$(f$_+$-1)(f$_-$-1). \textcolor{black}{Here the proximity to gelation measures how close the clusters are to having the capacity to potentially form an infinite cluster, i.e. proximity to percolation. The criterion for percolation for clusters with a cation-anion associated backbone has been previously derived to occur when 1=(f$_+$-1)(f$_-$-1)$p^*_{+-}p^*_{-+}$~\cite{mceldrew2020theory}, where $p^*_{ij}$ indicate the critical association probability for gelation. From this, we can define the proximity to gelation as 1-$\bar{p}_{+-}\bar{p}_{-+}$(f$_+$-1)(f$_-$-1) as when it approaches zero the system will approach gelation and as it becomes negative more of the system will be in the gel phase. }At most voltages especially large ones, the system is far from being a gel. At no applied field, the system is not close to the gel point; but as a positive potential is applied, the system gets closer and closer to forming a gel. At an applied potential of $\sim0.1~$V, the system almost reaches the gel point before the \textit{electric field induced associations in the EDL} are destroyed in favour of the saturation regime of anions at large potentials. There have been suggestions of similar effects in other electrolytes EDLs~\cite{Zhou2022Agg}. 

Finally, we describe the response of free ions and aggregates as seen in Fig.~\ref{fig:comp_001}\textcolor{black}{.d)}. As expected, the volume fraction of free alkali metal cations are practically 0 until the cation exchange has occurred at voltages more negative than -0.3$~$V. There are still many free anions at no applied voltage owing to the highly asymmetric mixture of the electrolyte. As positive potentials are applied, the volume fraction of free anions increases until $0.1$~V, when a plateau occurs. This plateau occurs at the same voltage as the bump in $\bar{p}_{-+}$, as the additional anions accumulating in the EDL are not free, but are associating to the additional alkali metal cations in the EDL. For larger potentials, the saturation regime of anions is reached, and the volume fraction of anions reaches 1. 

\begin{figure}[h!]
     \centering
     \includegraphics[width= 1\textwidth]{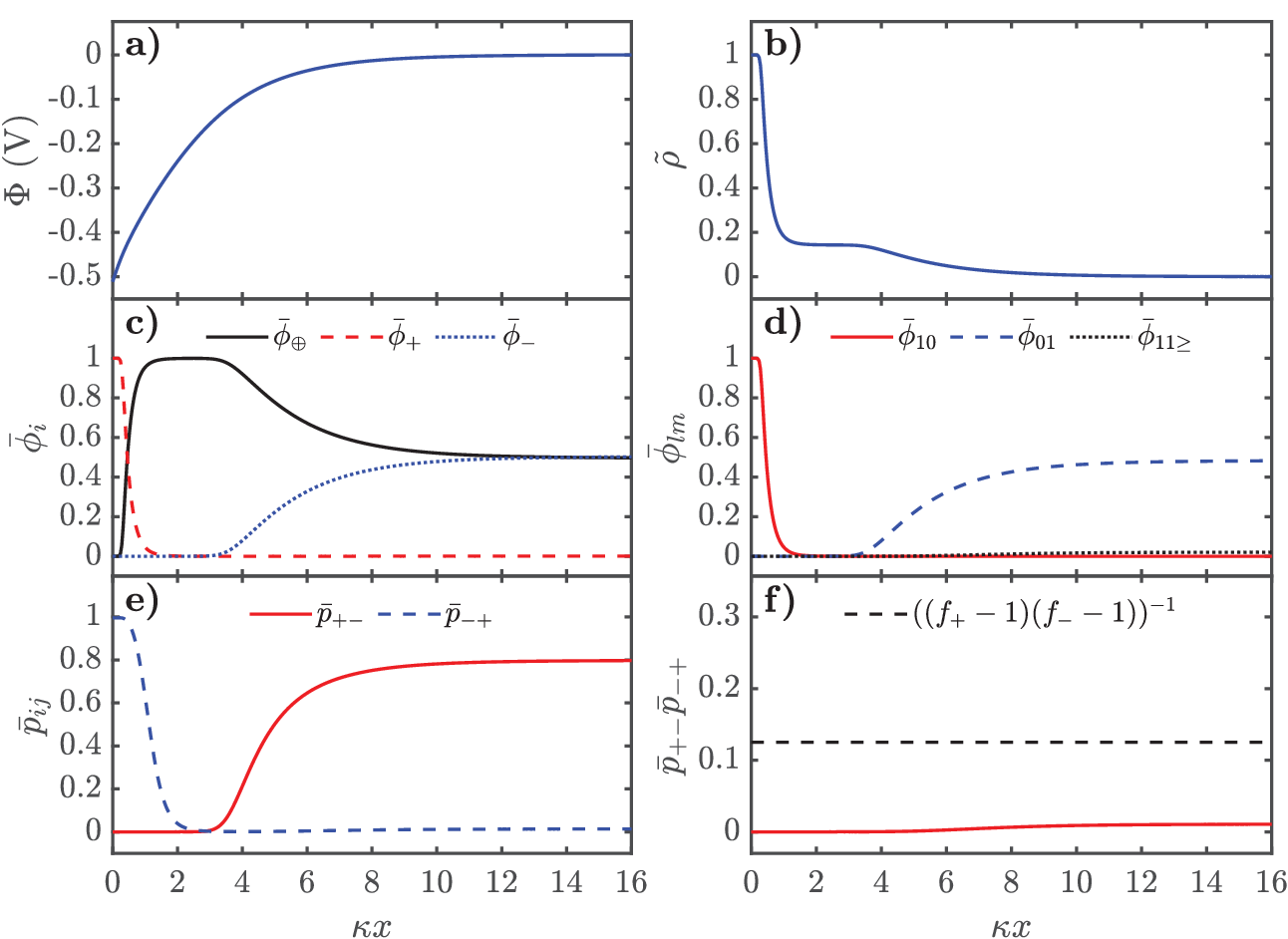}
     \caption{Distributions of properties of SiILs in the EDL as a function from the interface, in dimensionless units, where $\kappa$ is the inverse Debye length. \textcolor{black}{a)} Electrostatic potential\textcolor{black}{. b)} Dimensionless charge density, in units of $e/v_+$\textcolor{black}{. c)} Total volume fractions of each species\textcolor{black}{. d)} Volume fractions of free cations, free anions and aggregates\textcolor{black}{. e)} Association probabilities\textcolor{black}{. f)} Product of association probabilities, where the dashed line indicates the critical line. Here we use $x\textcolor{black}{_s} = 0.01$, $f_+=5$, $f_-=3$, $\xi_+=1$, $\xi_-=7$, $\xi_{\oplus}=7$, $\chi = -2$ k$_B$T, $\lambda_0=50$, $\epsilon_r=5$, and q$_s$ = -0.3 C/m$^2$.}
     \label{fig:neg}
\end{figure}

\begin{figure}[h!]
     \centering
     \includegraphics[width= 1\textwidth]{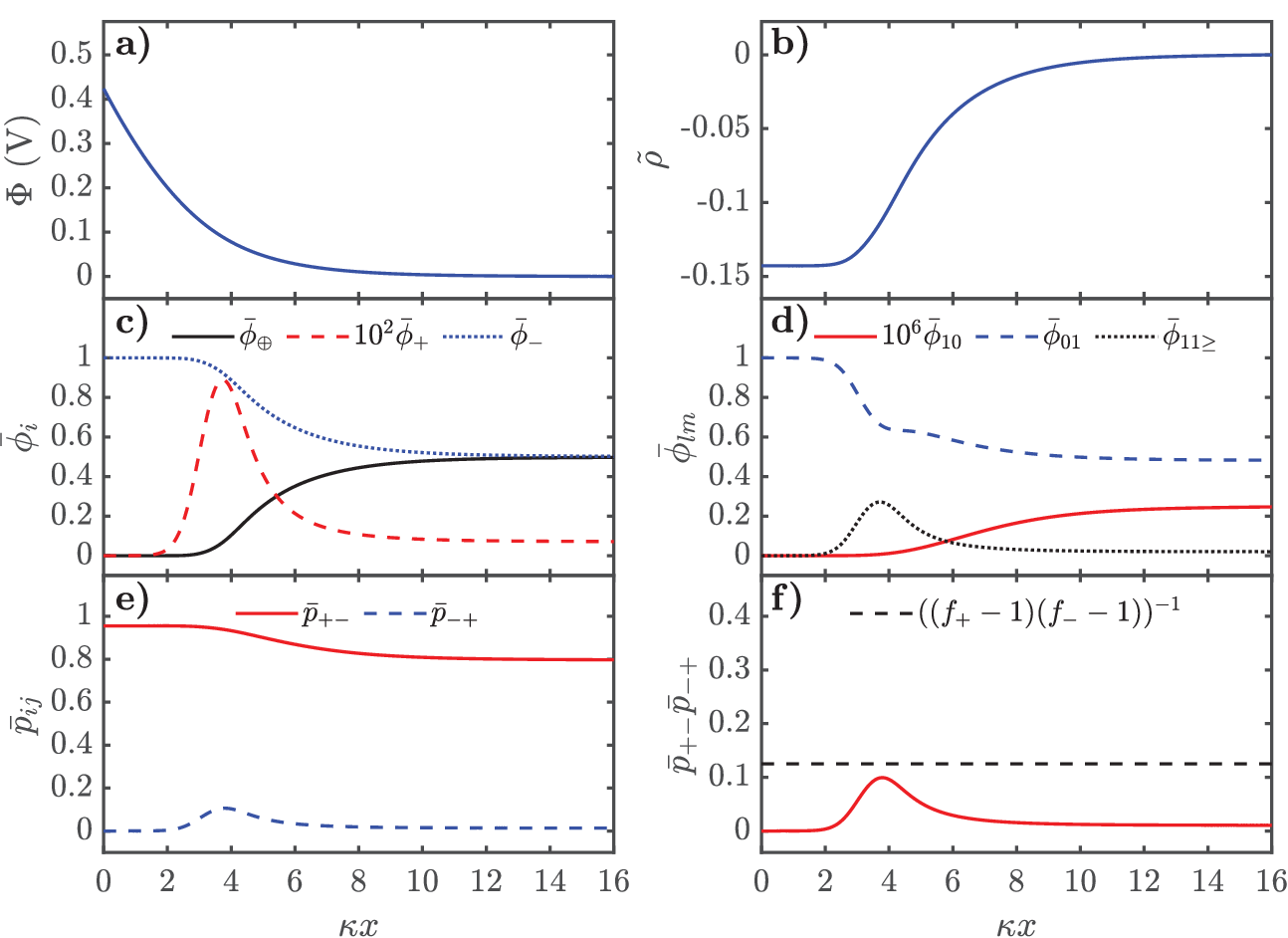}
     \caption{Distributions of properties of SiILs in the EDL as a function from the interface, in dimensionless units, where $\kappa$ is the inverse Debye length. \textcolor{black}{a)} Electrostatic potential\textcolor{black}{. b)} Dimensionless charge density, in units of $e/v_+$\textcolor{black}{. c)} Total volume fractions of each species\textcolor{black}{. d)} Volume fractions of free cations, free anions and aggregates\textcolor{black}{. e)} Association probabilities\textcolor{black}{. f)} Product of association probabilities, where the dashed line indicates the critical line. Here we use $x\textcolor{black}{_s} = 0.01$, $f_+=5$, $f_-=3$, $\xi_+=1$, $\xi_-=7$, $\xi_{\oplus}=7$, $\chi = -2$ k$_B$T, $\lambda_0=50$, $\epsilon_r=5$, and q$_s$ = 0.18 C/m$^2$.}
     \label{fig:pos}
\end{figure}

The description of the EDL properties displayed in Fig.~\ref{fig:comp_001} allows us to now understand the distributions of ions and association probabilities in the EDL. In Fig.~\ref{fig:neg}, we show properties of the EDL for negatively charged interfaces. In Fig.~\ref{fig:pos}, we display the similar figures but for positively charged interfaces. All plots are shown as a function from the electrified interface, in dimensionless units of distance, normalised by the inverse Debye length, $\kappa$, here 1/$\kappa$ is $\lambda_D \sim 0.3$ \AA\ for the EDL plots presented here. These plots were produced from numerically solving the grids shown in Fig.~\ref{fig:comp_001} in the Poisson equation, shown in Eq.~\eqref{PB}.

For negative voltages, the electrostatic potential decays in an exponential-like way to 0 in the bulk, as seen in Fig.~\ref{fig:neg}\textcolor{black}{.a)}. The charge density starts from 0 in the bulk, and first reaches a plateau of $\sim$1/7, in units of the number of lattice sites, which corresponds to the saturation regime of the IL cation, before the cation exchange occurs and the dimensionless charge density saturates at 1, i.e. the alkali metal cation saturates the EDL. As seen in Fig.~\ref{fig:neg}\textcolor{black}{.c)}, the cation exchange is now separated in space. Close to the interface, we find alkali metal cations are saturating the high field. After this dense layer of alkali metal cations, there is an additional saturation layer of the IL cations where the anions are also depleted. Naturally, there is a strong increase of $\bar{\phi}_{10}$ from the free cations right near the interface and a depletion of free anions $\bar{\phi}_{01}$ occurs further out from the interface. At negative potentials, there is only a monotonic destruction of aggregates, which can also be understood from $\bar{p}_{+-}\bar{p}_{-+}$, which can be used to quantify the gel transition, which also monotonically decreases. 

For positive voltages, the behaviour is more complex as seen in Fig.~\ref{fig:pos}. The electrostatic potential decays in a monotonic way from the interface and the charge density has the typical saturation curve of a Fermi-like function~\cite{Kornyshev2007,bazant2009a}. The structure of the electrolyte is complex despite these relatively simple appearing electrostatic potential and charge density distributions. As seen in Fig.~\ref{fig:pos}\textcolor{black}{.c)}, the total volume fraction of anions monotonically increases and the IL cation volume fraction monotonically decreases towards the interface. The alkali metal cations, however, have a strongly non-monotonic dependence, as shown in Fig.~\ref{fig:pos} where the volume fraction has been scaled up by a factor of 100 to make it more visible. Approaching the interface from the bulk, $\bar{\phi}_+$ dramatically increases, before reaching a peak, to then dramatically decrease in the saturation regime of the anions. Interestingly, as seen in Fig.~\ref{fig:pos}\textcolor{black}{.d)}, this increase in $\bar{\phi}_+$ exclusively corresponds to the formation of large aggregates in the EDL at moderate voltages, as can be seen from $\bar{\phi}_{10}$ monotonically decreasing towards the interface and a large peak in $\bar{\phi}_{11>}$. This large peak in $\bar{\phi}_{11>}$ also appears as a peak in $\bar{p}_{-+}$ in the EDL, as well as $\bar{p}_{-+}\bar{p}_{+-}$ approaching the critical gel line. Therefore, there are intermediate regions in space of the EDL at positive voltages where the SiIL becomes more aggregated than in the bulk. This finding is in contrast to previous theories of the EDL with associations, which only predicted the destruction of associations. \textcolor{black}{This finding comes generally from the asymmetry of the interactions, functionality, size and numbers of species found in SiIL. However, even with these asymmetries, there are specific compositions of SiIL and parameters such that the electric field induced associations are not present. The system will be most aggregated when the product of the association probabilities is maximized, which is not necessarily guaranteed to occur at zero voltage when asymmetries are present. The observation of electric field induced associations has been seen before in MD simulations of other electrolyte systems~\cite{ahmadiparidari2021enhancing,robin2021modeling, Zhou2022Agg}; a similar theory to the one presented here may be capable of explaining those observations.}

Overall, the EDL of SiILs is more complex than ILs. At negative voltages, the EDL has three regimes in space: (1) the bulk, where associations occur to create small negative aggregates (note this is for the $x\textcolor{black}{_s}$ chosen here, see Ref.~\citenum{McEldrewsalt2021} for the more general case); (2) moderate distances in the EDL, where the IL cation dominates; and (3) small distances a saturation layer of alkali metal cations. Similarly for positive voltages, there are three regimes in space: (1) the bulk, where associations occur to create small negative aggregates; (2) moderate distances in the EDL, where the change in composition occurs large aggregates to appear from electric field induced associations; and (3) the saturation regime close to the interface. This is analogous to the overscreening-saturation transition that is known to occur in ILs~\cite{Bazant2011,Pedro2020,Goodwin2021Rev}, and suggests that in SiILs it might be accompanied by dramatic changes in the associations. Moreover, the link between associations and overscreening was established in Ref.~\citenum{avni2020charge}, and further discussed in Ref.~\citenum{Goodwin2022EDL}. These associations occur without any visible feature in the charge density, and therefore, must be probed by computing the associations in the EDL through more sophisticated theories~\cite{avni2020charge} or through experimental means. 

\begin{figure}[h!]
 \centering
 \includegraphics[width= 1\textwidth]{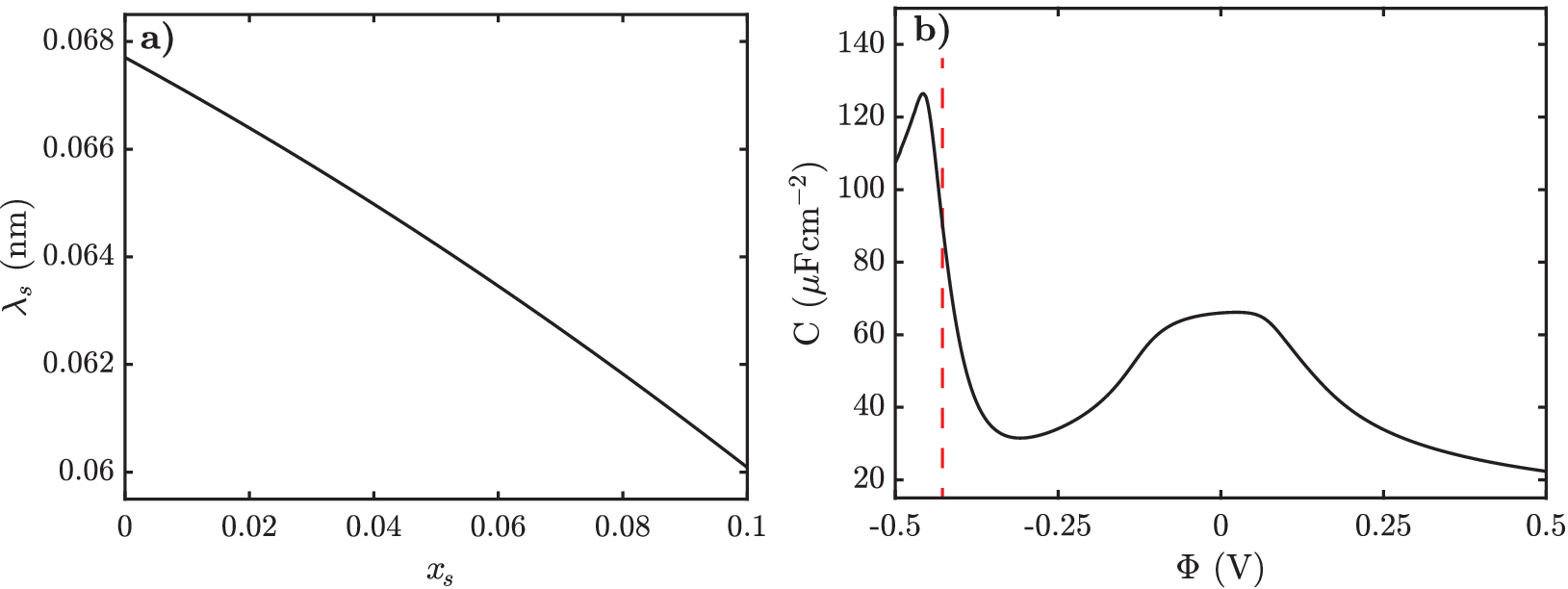}
 \caption{\textcolor{black}{a)} Screening length of SiIL as a function of mole fraction of alkali metal salt\textcolor{black}{. b)} Differential capacitance of SiIL with $x\textcolor{black}{_s} = 0.01$, as a function of electrostatic potential. Here we use $f_+=5$, $f_-=3$, $\xi_+=1$, $\xi_-=7$, $\xi_{\oplus}=7$, $\chi = -2$ k$_B$T, $\lambda_0=50$, and $\epsilon_r=5$. \textcolor{black}{b.) Here the red dashed line at -0.43 V indicates the cation exchange voltage.}}
 \label{fig:screencap}
\end{figure}

Next we turn to how the screening length, $\lambda_s$, varies with the mole fraction of alkali metal salt, $x\textcolor{black}{_s}$, in the pre-gel regime. The screening length was extracted from fitting an exponential function to the electrostatic potential at small potentials for various $x\textcolor{black}{_s}$. In Fig.~\ref{fig:screencap}\textcolor{black}{.a)}, we observe that as $x\textcolor{black}{_s}$ increases the screening length decreases. It is typically expected, however, that the screening length increases as the electrolyte becomes more associated, and as $x\textcolor{black}{_s}$ increases the SiILs becomes more associated. Here we find that the screening length actually decreases with $x\textcolor{black}{_s}$ because of the formation of small, highly charged aggregates, which is the opposite of conventional intuition. 

Lastly, we predict the differential capacitance for SiILs as a function of potential, as shown in Fig.~\ref{fig:screencap}\textcolor{black}{.b)}. At low potentials the differential capacitance takes on a slightly asymmetry bell shape~\cite{Kornyshev2007}. At large positive potentials, we observe the typical decay in differential capacitance~\cite{Kornyshev2007,kilic2007a}. At negative potentials, however, the differential capacitance initially decays as expected from the saturation of the IL cations~\cite{Kornyshev2007,kilic2007a}. However, at a critical potential we observe a large peak in differential capacitance which can be attributed to the cation exchange shortening the EDL. This satellite peak can be used to verify the predictions of our theory.

In the Supplemental Material, we display additional plots and discuss the role of changing parameter and variable values from the values chosen in the main text. 

\section{Discussion}

In Ref.~\citenum{Haskins2016}, Haskins \textit{et al.} performed molecular dynamics simulations of Li[TFSI] in [pyr14][TFSI] at electrified interfaces. For positive voltages, they found that the anions dominate the first layer of ions with some IL cations. After this layer, there is a sharp peak in alkali metal cations residing between layers of anions~\cite{Haskins2016}. The second peak of anions does not have an associated second peak of Li$^+$ ions, which could indicate that the Li$^+$ ions in between these peaks are associating to anions in multiple layers of the overscreening structure~\cite{Haskins2016}. Whereas at negative potentials, the IL cation dominates the first layer of the EDL, with no Li$^+$ ions~\cite{Haskins2016}. In the second layer of ions, there is pronounced overscreening with an increase in the number density of anions accompanied by a dramatic increase in Li$^+$ ions. In this second layer, large aggregates could be forming. These MD results are promising in confirming the possible electric field induced associations, but further MD simulations must be done to confirm the predictions of our theory. The EDL of SiILs is an extremely understudied problem and we hope our theory will provide a framework in which it can be understood.

A few studies have looked at the differential capacitance of SiILs~\cite{yamagata2013charge,Haskins2016}. In Ref.~\citenum{yamagata2013charge}, Yamagata \textit{et al.} experimentally measured the differential capacitance of Li[TFSI] in [EMIM][FSI] and Li[TFSI] in [EMIM][TFSI] at negative potentials. In their experiments, they observe a significant difference between these two SiILs differential capacitance at large negative potentials in comparison to the neat ILs. For Li[TFSI] in [EMIM][FSI], they observe that at low potential the differential capacitance is lower than the neat IL; however for larger negative potentials, the SiILs differential capacitance grows significantly. The sharp rise in differential capacitance seen in the Li[FSI] in [EMIM][FSI] could be coming from a cation exchange occurring in the system. They proposed a schematic of their EDL for this system where the Li$^+$ ion is the cation present at the interface, while still being associated with an FSI anion. In our theory, this peak in the differential capacitance comes from the cation exchange, which is a similar origin to that suggested by Yamagata \textit{et al.}~\cite{yamagata2013charge}. In contrast in the Li[TFSI] in [EMIM][TFSI] system, they do not see this differential capacitance growth at large negative potentials. They consider the EDL for this case to have the interface occupied by the EMIM cations and, therefore, has not undergone the cation exchange yet. This schematic is similar to what is predicted from the MD results in Ref.~\citenum{Haskins2016}, as discussed above. For certain parameters of our theory, we do not obtain this satellite peak from the cation exchange as shown in the Supplemental Material.

Experimentally, it has been found that ILs and other concentrated electrolytes have extremely long force decay lengths as measured from surface force measurements~\cite{Gebbie2013,Marzal2014,Gebbie2015,Smith2016,Han2020,Han2021WiSE}. The assumption of Gebbie \textit{et al.}~\cite{Gebbie2013} was that these forces were electrostatic in origin, owing to a large renormalisation in the concentration of free charge carriers. While this topic remains controversial and unresolved~\cite{feng2019free,jones2021bayesian,krucker2021underscreening,Espinosa2023rev}, we would like to briefly highlight that the aggregation of the alkali metal cations and anions in SiILs was able to successfully explain the transport properties in these super-concentrated electrolytes~\cite{molinari2019general,molinari2019transport,McEldrewsalt2021}. Therefore, aggregates could shed some light onto these measurements of surface force interactions in SiILs and long ranged interactions in other concentrated electrolytes. \textcolor{black}{Recently, experimental evidence for the connection between the cluster-like (soft) structure in a SiIL composed of NaTFSI and EMIM TFSI, at various mole ratios, and long-range (non-exponentially decaying) surface forces measured with a Surface Forces Apparatus~\cite{Zhang2024}. Furthermore, the long-ranged interactions point at the presence of aggregates with a high aspect ratio at the mica interfaces that contribute to the surface force, rather than a purely electrostatic origin. We believe that the theory described here could shed some light not only into these measurements, but also into the long ranged interactions in other concentrated electrolytes~\cite{Gebbie2013,Marzal2014,Gebbie2015,Smith2016,Han2020,Han2021WiSE}.}
 
We have chosen not to include the $\alpha$-parameter introduced in Ref.~\citenum{goodwin2017mean} and adopted in Ref.~\citenum{Goodwin2022EDL}. Therefore, the voltages in our curves are smaller than what is expected from MD simulations and experiments which should be kept in mind when interpreting the results. As outlined in Ref.~\citenum{Goodwin2022EDL}, there are also some conceptual and technical limitations with our presented theory. Firstly, overscreening, the alternating layers of cations and anions, is represented through the thermo-reversible associations between the ions; but, the internal structure of these aggregates is not explicitly defined here. Therefore, we cannot explicitly obtain decaying oscillations in the charge density but need to recognise that overscreening is included through these associations. In SiILs, the electric field induced associations at positive voltages further complicates this problem. If the potential is oscillating, there will be positive potential values at both the cathode and anode. Thus, we might, in fact, predict that there are electric field induced associations at both charged interfaces. This will need to be further studied through MD simulations. In addition, our theory can describe well the diffuse EDL, but not the interfacial layer of ions, where a breakdown of the cluster distribution should occur and specific interactions with the electrode could dominate. This latter point could be predicted by the MD simulations of Haskins \textit{et al.}~\cite{Haskins2016} which found that the IL cation prefers to reside in the first layer of ions, where we expect the alkali metal cations to dominate the saturation regime. However, this could also be a problem with short equilibration times of the simulations as discussed later. 

The largest limitation of our theory is that we find the screening length is less than 1 \AA. Since $\lambda_s$ is smaller than an individual alkali metal cation ($\sim 2.8$ \AA, as used in this work), and much less than the size of the small clusters predicted to be in the SiIL solution~\cite{bazant2009a}. This points to the theory not being self-consistent in terms of its electrostatic predictions, as is generally known to be a problem of mean-field lattice gas models which is discussed in detail in Ref.~\citenum{bazant2009a}. This common failing of mean-field theories can be corrected with more involved modified Poisson-Boltzmann equations that can account for higher-order correlations and/or non-local effects. This issue can be addressed to some extent in local mean field models such as by accounting for higher order local terms such as done in BSK theory~\cite{Bazant2011} or via modifying the Coulomb interactions~\cite{adar2019screening}. However typically to overcome this inconsistency, one must turn to a non-local model that can capture the entropic effects of excluded volume in a significantly better fashion. The issue from the current inconsistency indicates caution should be used when comparing the model's predictions of the spatial profile of the ions and clusters near the surface~\cite{bazant2009a}. Importantly even with this inconsistency, mean-field models can have valuable predictive power in integrated quantities such as the capacitance of the double layer, the general trends in what accumulates in the EDL, and importantly, the new predictions for how association probabilities change within the EDL.

Finally, we have presented an analysis of the equilibrium properties of SiILs. One of the noteworthy properties of these systems is their transport properties. As outlined in Ref.~\citenum{Goodwin2023}, the dynamics of this model can also be investigated, which could yield some interesting effects. Moreover as this system has an active cation for intercalation, developing the theory of coupled ion electron transfer reactions to include microscopic solvation effects could be a promising place to start~\cite{Fraggedakis2021,bazant2023unified}.

\section{Conclusions}

Here we have developed a theory for the electrical double layer (EDL) of salt-in-ionic liquids (SiILs). These electrolytes are interesting for applications in energy storage but their behavior at electrified interfaces is not well studied. Here, we found a highly asymmetric response of the SiILs in the EDL. At negative potentials, associations are destroyed in the EDL; but, a cation exchange can occur when the IL cation initially saturates in the EDL but is later replaced by the smaller, associating alkali metal cation. In contrast, the positive potentials predict there to be electric field induced associations, which may lead to gelation. At large positive voltages we obtain the saturation regime for anions. All of these EDL features appear to be captured in the predicted differential capacitance profile for SiILs. Overall, we hope that the predictions made by this theory will inspire further work from molecular dynamics simulations to understand these SiILs, or for experimental measurements to be performed to probe the cation exchange or electric field induced associations.

\section{Acknowledgements}

D.M.M. \& M.Z.B. acknowledge support from the Center for Enhanced Nanofluidic Transport \textcolor{black}{2} (CENT$^{\textcolor{black}{2}}$), an Energy Frontier Research Center funded by the U.S. Department of Energy (DOE), Office of Science, Basic Energy Sciences (BES), under award \# DE-SC0019112. D.M.M. also acknowledges support from the National Science Foundation Graduate Research Fellowship under Grant No. 2141064. M.M. and M.Z.B. acknowledge support from an Amar G. Bose Research Grant. We gratefully acknowledge financial support from the National Science Foundation under grants DMR-1904681 and CBET-1916609 to R.M.E.M.

\renewcommand{\theequation}{S\arabic{equation}} 
\renewcommand{\thefigure}{S\arabic{figure}}  
\setcounter{figure}{0} 
\setcounter{equation}{0}   

\begin{center}
    \section*{Supplemental Material}
\end{center}

\section{Model Predictions for Alternative Parameters}

In the main text, the analysis centered around parameters that were extracted from MD simulations and in \textcolor{black}{the pre-gel regime} where we could access its predictions~\cite{McEldrewsalt2021}. Extending this model into the post-gel regime follows a similar framework as with previous works, with the additional change of expanding upon the regular solution interaction term to be capable of capturing the effects of having gel-species~\cite{mceldrew2020theory,McEldrewsalt2021}. A sample of alternative parameter choices are presented here to better demonstrate the underlying physics that this model may predict and discuss briefly the basis behind why the various changes in electrolyte behavior arise.

\begin{figure}[h!]
     \centering
     \includegraphics[width= 1\textwidth]{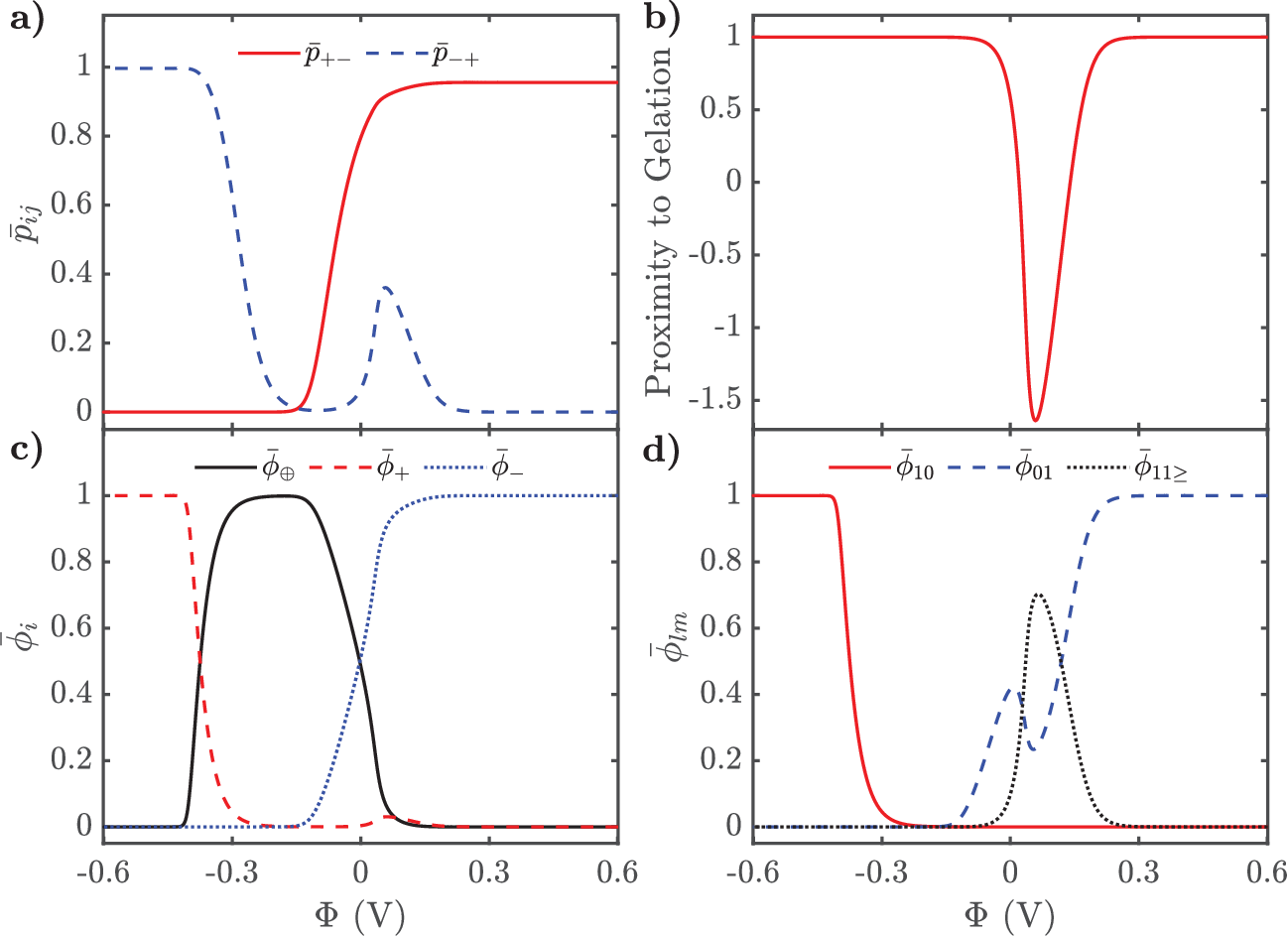}
     \caption{Properties of the EDL of SiILs at a \underline{higher mole fraction of alkali metal salt} as a function of applied electrostatic potential. \textcolor{black}{a)} Association probabilities\textcolor{black}{. b)} Proximity to gelation, $1-\bar{p}_{+-}\bar{p}_{-+}(f_+-1)(f_--1)$\textcolor{black}{. c)} Total volume fractions of each species\textcolor{black}{. d)} Volume fraction of free cations, anions and clusters. Here we use $x\textcolor{black}{_s} = 0.05$, $f_+=5$, $f_-=3$, $\xi_+=1$, $\xi_-=7$, $\xi_{\oplus}=7$, $\chi = -2$ k$_B$T, and $\lambda_0=50$. Here we observe that sufficiently high mole fractions of alkali metal salt lead to a local minimum in the free IL anions emerging around low positive potentials. Besides the formation of a local minimum from the increased presence of alkali metal salt, the predictions of this electrolytes composition in the EDL does not depart drastically from the case discussed in the main paper. This increased presence also leads to a decrease in the voltage plateau before the cation exchange occurs. The influence of gelation is ignored naively for the current study; therefore, more nuanced differences may result if the influence of gelation is significant. Note we observe gel from 0.020 V to 0.140 V.}
     \label{fig:SI_comp_001}
\end{figure}

\begin{figure}[h!]
     \centering
     \includegraphics[width= 1\textwidth]{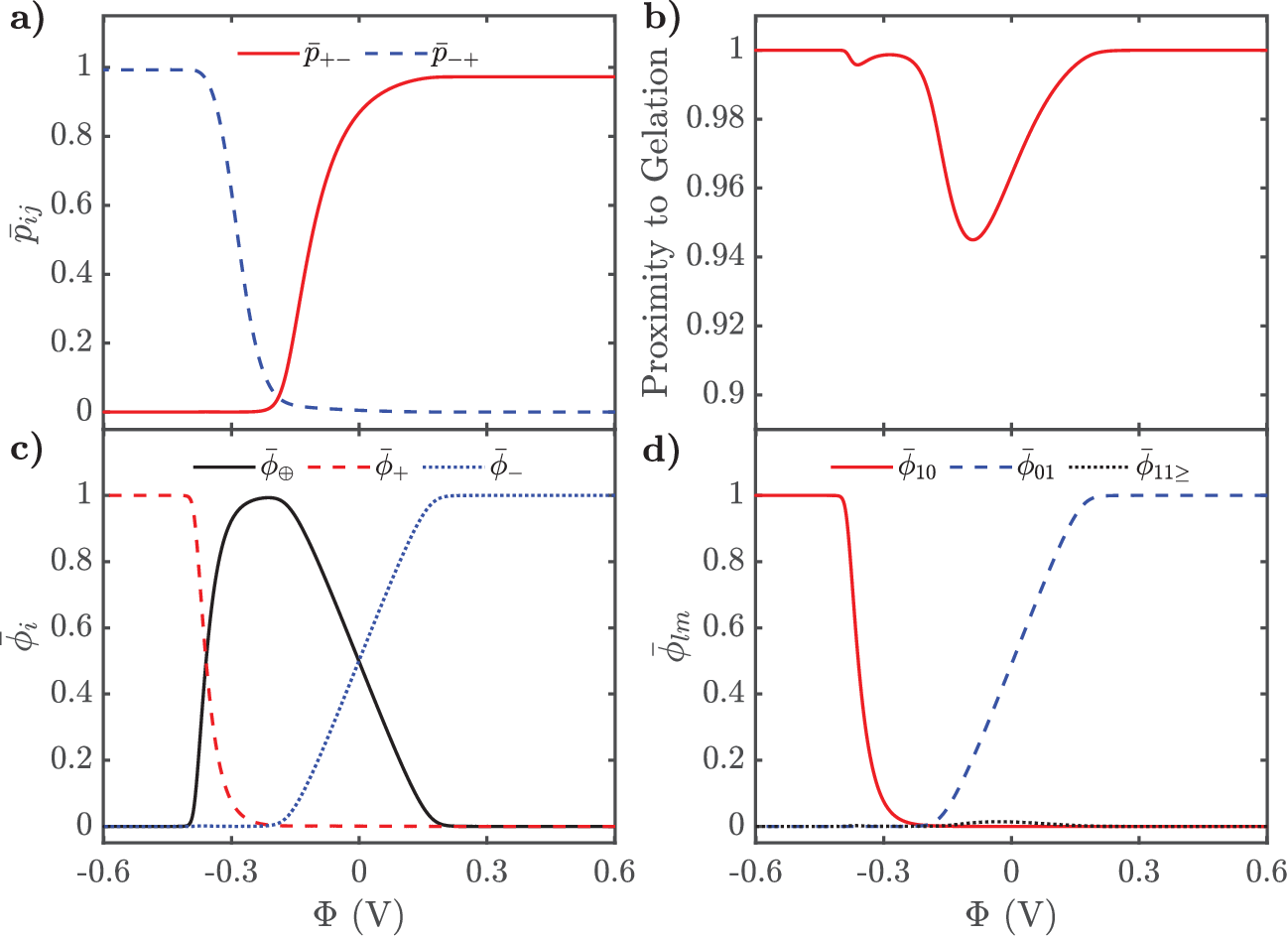}
     \caption{Properties of the EDL of SiILs with \underline{flipped functionalities} as a function of applied electrostatic potential. \textcolor{black}{a)} Association probabilities\textcolor{black}{. b)} Proximity to gelation, $1-\bar{p}_{+-}\bar{p}_{-+}(f_+-1)(f_--1)$\textcolor{black}{. c)} Total volume fractions of each species\textcolor{black}{. d)} Volume fraction of free cations, anions and clusters. Here we use $x\textcolor{black}{_s} = 0.01$, $f_+=3$, $f_-=5$, $\xi_+=1$, $\xi_-=7$, $\xi_{\oplus}=7$, $\chi = -2$ k$_B$T, and $\lambda_0=50$. Here we see a deviation in the predictions of the SiILs properties compared to the main case. The decrease of the bulk's proximity to gelation is expected from the composition of the SiIL. From this decrease, the increase in aggregation is seen at modest applied negative potentials. Beyond this, we observe the voltage plateau shrinks. We note \textcolor{black}{that} a secondary increase in aggregation occurs after the voltage plateau. This effect helps highlight the role of the regular solution interactions in the fluctuation of aggregation in the EDL. This change leads to the enhancement of aggregation occurring at a modest applied negative potentials, as well as secondary enhancement in aggregation emerging at higher but not extreme applied negative potentials. The nuanced interplay of the SiIL parameters can lead to the secondary enhancement being larger than the first enhancement as well as shifting the enhancements \textcolor{black}{to different voltages}.}
     \label{fig:SI_comp_002}
\end{figure}

\begin{figure}[h!]
     \centering
     \includegraphics[width= 1\textwidth]{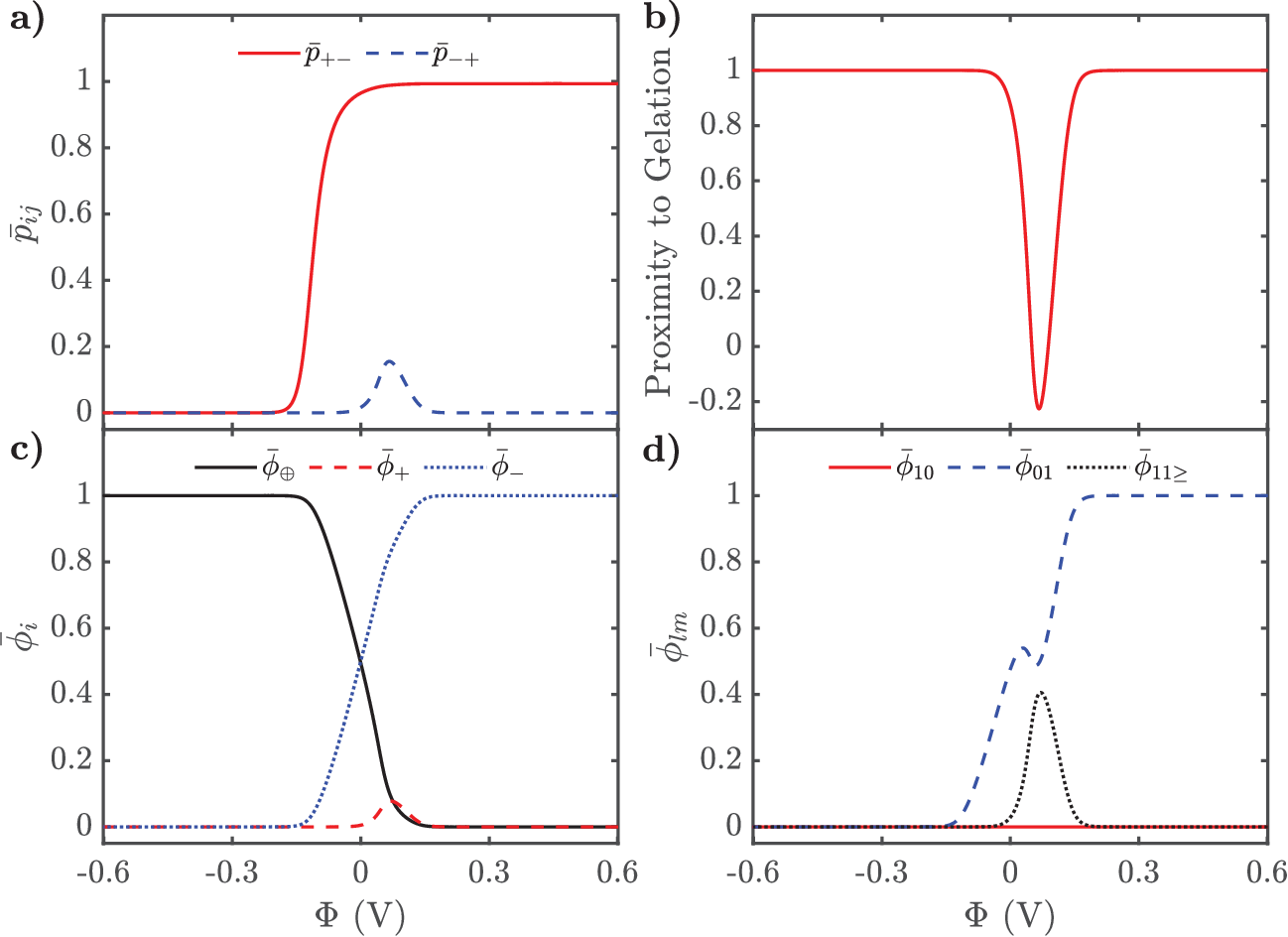}
     \caption{Properties of the EDL of SiILs with \underline{equal sized species} as a function of applied electrostatic potential. \textcolor{black}{a)} Association probabilities\textcolor{black}{. b)} Proximity to gelation, $1-\bar{p}_{+-}\bar{p}_{-+}(f_+-1)(f_--1)$\textcolor{black}{. c)} Total volume fractions of each species\textcolor{black}{. d)} Volume fraction of free cations, anions and clusters. Here we use $x\textcolor{black}{_s} = 0.01$, $f_+=5$, $f_-=3$, $\xi_+=\xi_-=\xi_{\oplus}=1$, $\chi = -2$ k$_B$T, and $\lambda_0=50$. We observe that the bulk system initially is closer to gelation and obtains a similar enhancement at small positive potentials as before. The system here enters the post-gel regime; however, more nuanced effects may not be captured as previously discussed. Additionally unlike in the main case, we do not see the cation exchange occurring in our system even for large negative potentials. This finding is expected as the alkali metal cation no longer has a larger charge density. Lastly, we can note the alkali metal cation enhancement \textcolor{black}{as} seen in \textcolor{black}{c)}, which is harder to observe in the main case, given the significant size difference between the species. Note we observe gel from 0.050 V to 0.088 V.}
     \label{fig:SI_comp_003}
\end{figure}

\begin{figure}[h!]
     \centering
     \includegraphics[width= 1\textwidth]{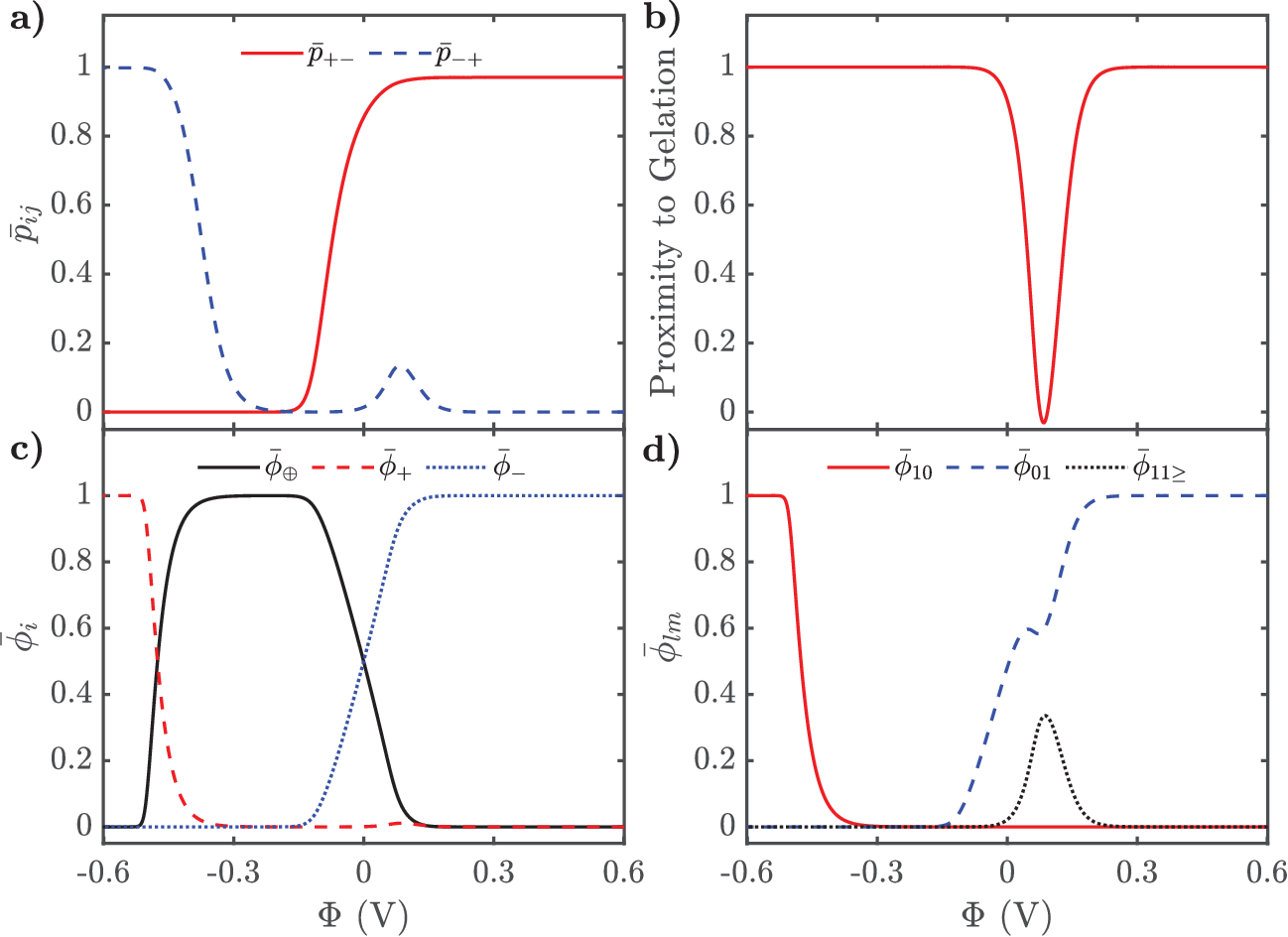}
     \caption{Properties of the EDL of SiILs with a \underline{stronger ``bare" association constant, $\lambda_0$,} as a function of applied electrostatic potential. \textcolor{black}{a)} Association probabilities\textcolor{black}{. b)} Proximity to gelation, $1-\bar{p}_{+-}\bar{p}_{-+}(f_+-1)(f_--1)$\textcolor{black}{. c)} Total volume fractions of each species\textcolor{black}{. d)} Volume fraction of free cations, anions and clusters. We use $x\textcolor{black}{_s} = 0.01$, $f_+=5$, $f_-=3$, $\xi_+=1$, $\xi_-=7$, $\xi_{\oplus}=7$, $\chi = -2$ k$_B$T, and $\lambda_0=75$. Here we see that the bulk gets closer to gelation. This \textcolor{black}{change} leads to a similar prediction in regards to the enhancement of aggregations at modest applied positive potentials that is seen when the mole fraction of alkali metal salt was increased. Uniquely here, the voltage plateau expands, which differs from the voltage plateau contraction seen with an increased mole fraction of alkali metal salt. This expansion comes primarily from decreasing the favorability of the associating species being brought into the EDL. Note we observe gel from 0.076 V to 0.093 V.}
     \label{fig:SI_comp_004}
\end{figure}

\begin{figure}[h!]
     \centering
     \includegraphics[width= 1\textwidth]{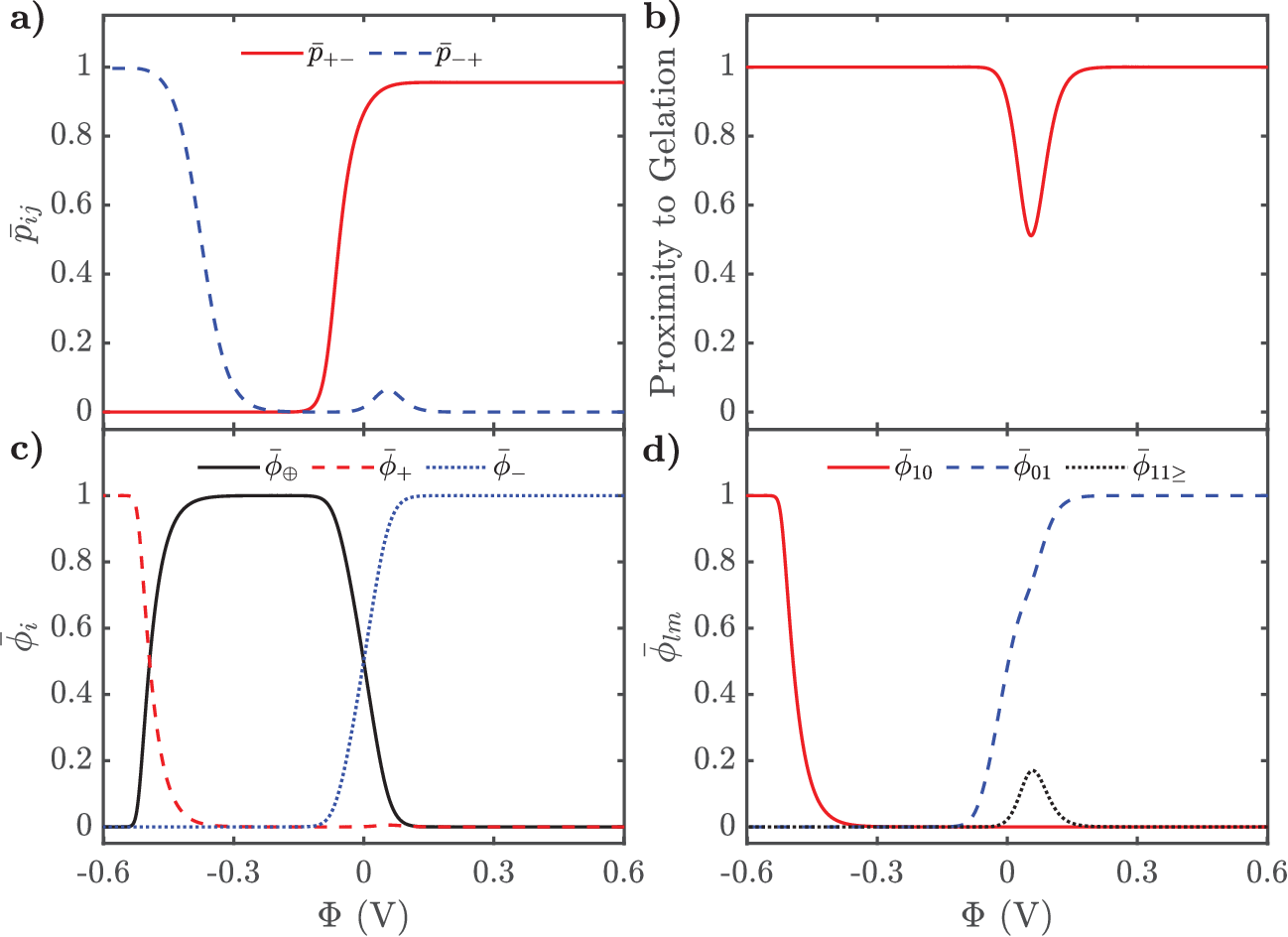}
     \caption{Properties of the EDL of SiILs with a \underline{weaker regular solution interaction strength, $\chi$,} as a function of applied electrostatic potential. \textcolor{black}{a)} Association probabilities\textcolor{black}{. b)} Proximity to gelation, $1-\bar{p}_{+-}\bar{p}_{-+}(f_+-1)(f_--1)$\textcolor{black}{. c)} Total volume fractions of each species\textcolor{black}{. d)} Volume fraction of free cations, anions and clusters. Here we use $x\textcolor{black}{_s} = 0.01$, $f_+=5$, $f_-=3$, $\xi_+=1$, $\xi_-=7$, $\xi_{\oplus}=7$, $\chi = -1$ k$_B$T, and $\lambda_0=50$. Here the enhancement of aggregates at low positive potentials decreases. This weaker regular solution interaction strength also leads to an expanded voltage plateau before the cation exchange occurs, as compared to the main case. This effect comes from the impact of having weaker regular solution interactions, meaning it's less able to compensate for the unfavorability of bringing the associating species into the EDL.}
     \label{fig:SI_comp_005}
\end{figure}

\begin{figure}[h!]
     \centering
     \includegraphics[width= 1\textwidth]{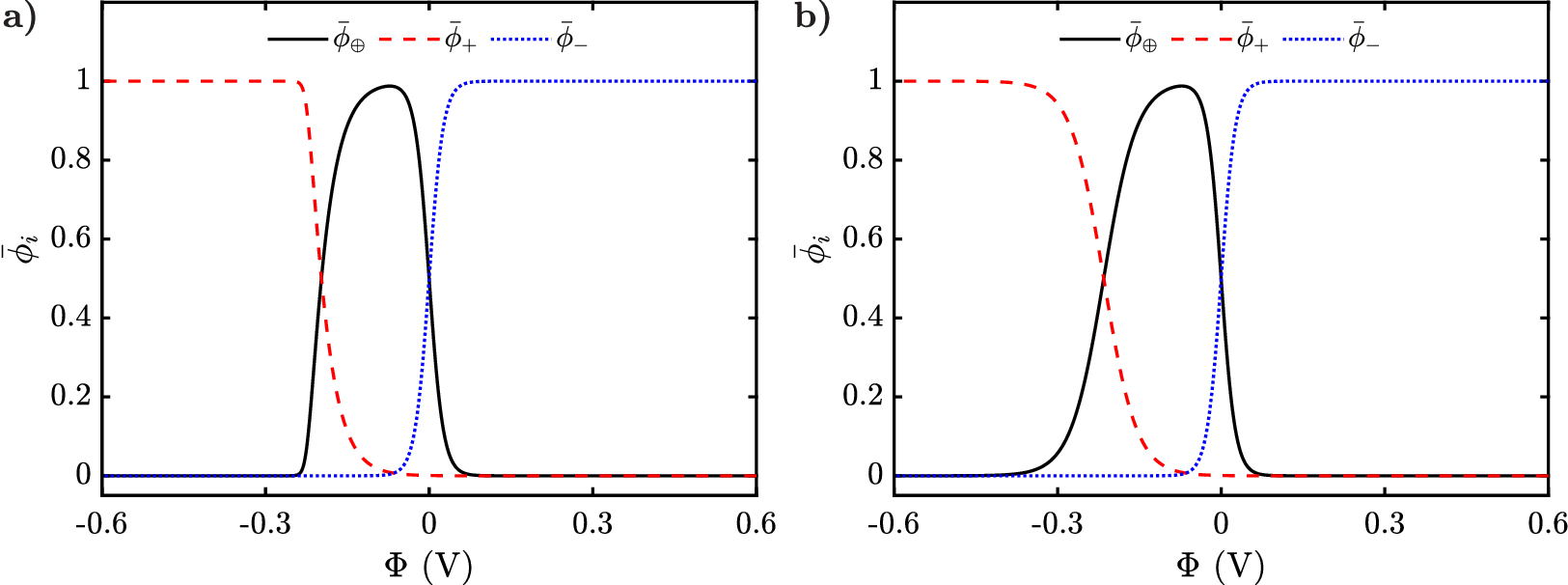}
     \caption{\textcolor{black}{Total volume fractions of each species in EDL of SiILs without associations and regular solution interactions as a function of applied electrostatic potential. a) Incompressible model. b) Asymmetric 3-component Langmuir model~\cite{mceldrew2018}. Here we use $x_s = 0.01$, $\xi_+=1$, $\xi_-=7$, and $\xi_{\oplus}=7$ for both models. Here one can observe the cation exchange at lower electric potentials than in the main case, occurring at -0.20 V for the Incompressible model and -0.22 V for the Asymmetric 3-component Langmuir model. This exchange is driven by the difference in the charge density between the cations as well as the bulk composition of the electrolyte solution.}}
     \label{fig:SI_Filling}
\end{figure}

\begin{figure}[h!]
     \centering
     \includegraphics[width= 1\textwidth]{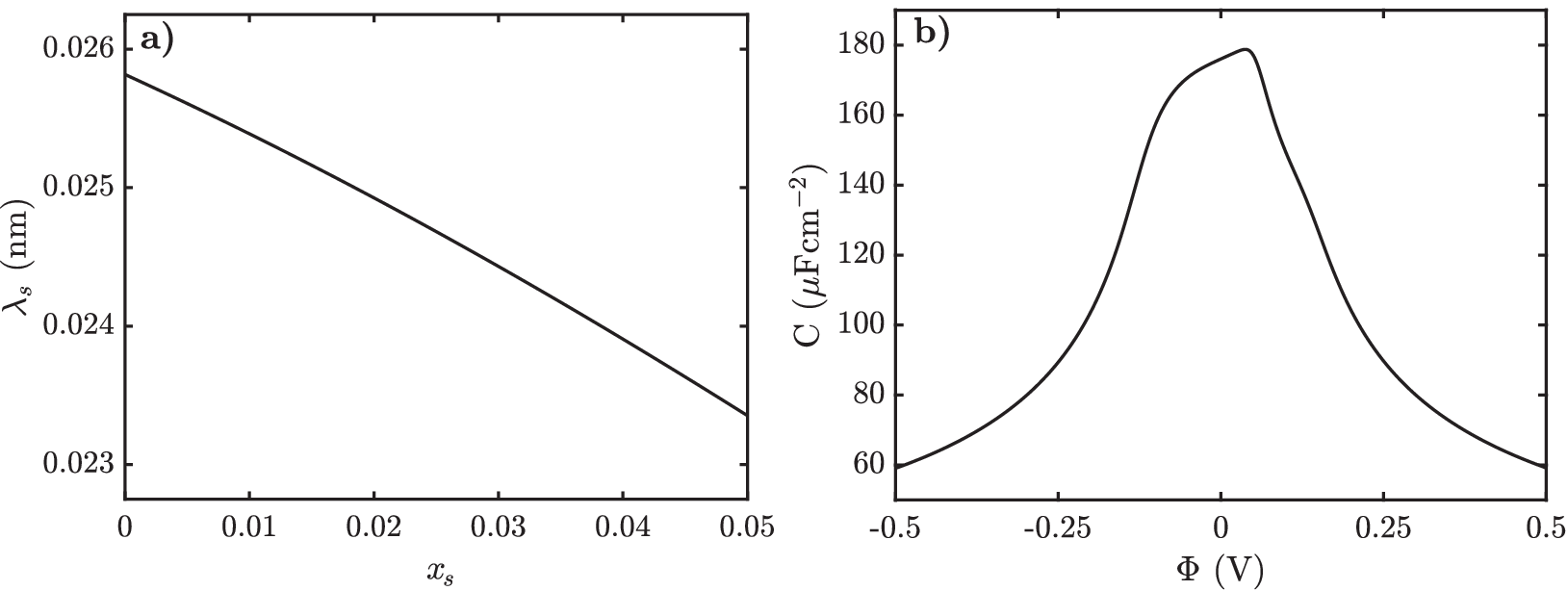}
     \caption{\textcolor{black}{a)} Screening length of SiIL with \underline{equal sized species} as a function of the mole fraction of alkali metal salt\textcolor{black}{. b)} Differential capacitance of SiIL with \underline{equal sized species} and $x\textcolor{black}{_s} = 0.01$, as a function of electrostatic potential. Here we use $f_+=5$, $f_-=3$, $\xi_+$=$\xi_-$=$\xi_{\oplus}$=1, $\chi = -2$ k$_B$T, $\lambda_0=50$, and $\epsilon_r=5$. \textcolor{black}{a)} We see here that the screening length is decreased significantly compared to the case seen in the main paper. This drop across the mole fraction of alkali metal salt points towards the dominant role the formation of aggregates have on the screening length. \textcolor{black}{b)} We can note that the satellite peak associated with the cation exchange disappears. This change follows from the fact that there is now no difference in the charge density of our cations which leads to no cation exchange. Here, we can also see more prominently the effects of the induced associations at modest positive potentials via the low potential differential capacitance peak being even larger. Additionally, we observe that at low magnitudes of potential that the differential capacitance decays asymmetrically before returning to similar rates of decay at larger magnitudes of potential. After 0.050 V we would expect gel to be present in some parts of the EDL.}
     \label{fig:SI_screencap}
\end{figure}

\begin{figure}[h!]
     \centering
     \includegraphics[width= .7\textwidth]{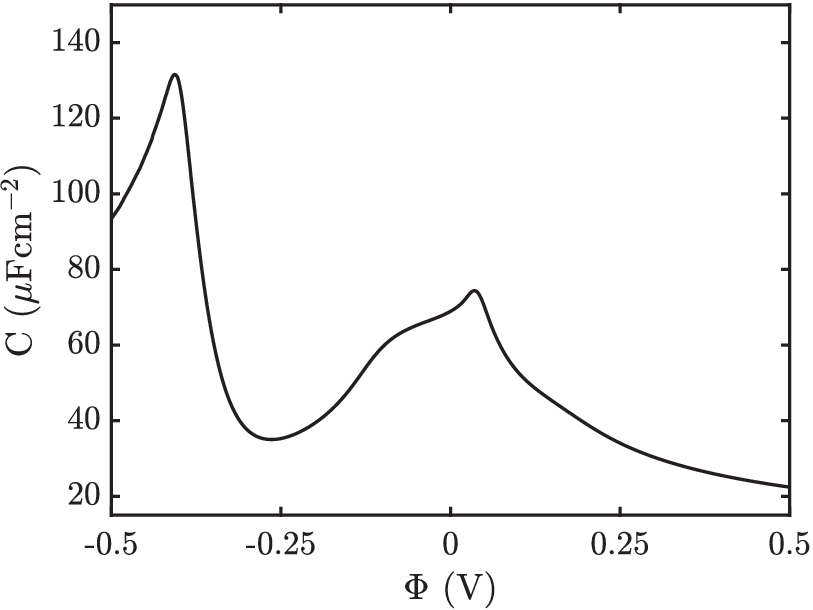}
     \caption{Differential capacitance of SiIL at a \underline{higher mole fraction of alkali metal salt} as a function of electrostatic potential. Here we use $x\textcolor{black}{_s} = 0.05$, $f_+=5$, $f_-=3$, $\xi_+=1$, $\xi_-=7$, $\xi_{\oplus}=7$, $\chi = -2$ k$_B$T, $\lambda_0=50$, and $\epsilon_r=5$. Here we can observe that the induced associations are even more enhanced at modest positive potentials, which appears to make the low potential differential capacitance peak even larger. Additionally, we can note as potentials increase that the decay is asymmetric and steeper initially for the positive potentials. Once again, we see the secondary satellite peak associated with the cation exchange. After 0.020 V we would expect gel to be present in some parts of the EDL.}
     \label{fig:SI_cap_001}
\end{figure}

\begin{figure}[h!]
     \centering
     \includegraphics[width= .7\textwidth]{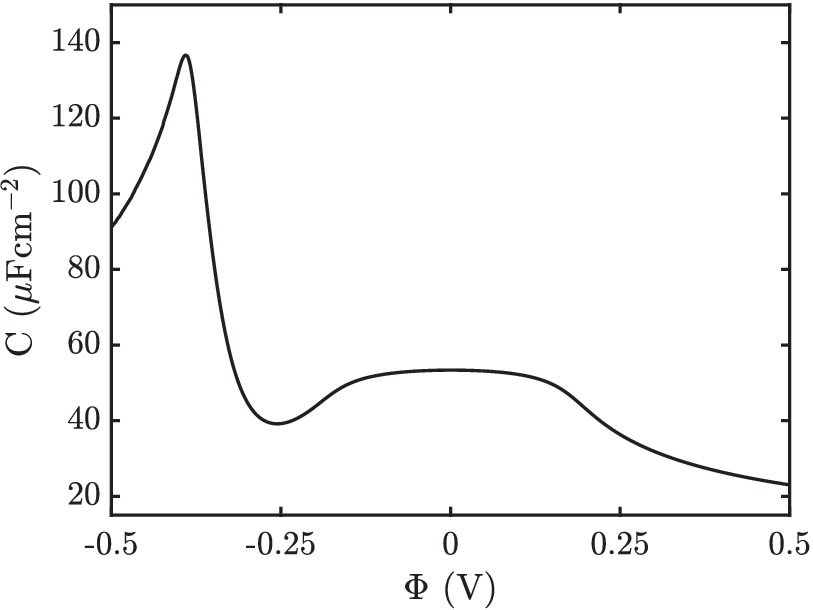}
     \caption{Differential capacitance of SiIL with \underline{flipped functionalities} as a function of electrostatic potential. Here we use $x\textcolor{black}{_s} = 0.01$, $f_+=3$, $f_-=5$, $\xi_+=1$, $\xi_-=7$, $\xi_{\oplus}=7$, $\chi = -2$ k$_B$T, $\lambda_0=50$, and $\epsilon_r=5$. Here we can observe that the initial bell shape is elongated. This influence may be due to the induced associations now occurring at modest negative potentials. Once again, we note the secondary satellite peak associated with the cation exchange.}
     \label{fig:SI_cap_002}
\end{figure}

\begin{figure}[h!]
     \centering
     \includegraphics[width= .7\textwidth]{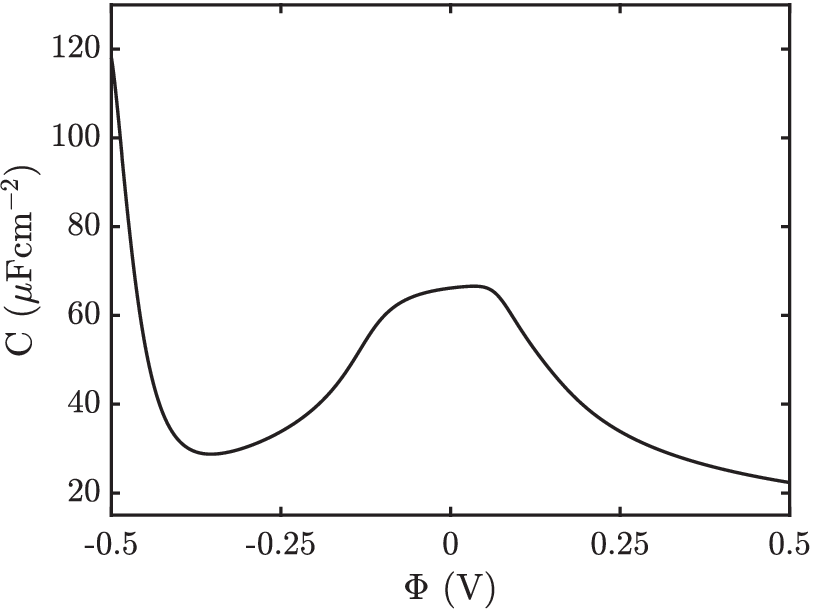}
     \caption{Differential capacitance of SiIL with a \underline{stronger ``bare" association constant, $\lambda_0$,} as a function of electrostatic potential. Here we use $x\textcolor{black}{_s} = 0.01$, $f_+=5$, $f_-=3$, $\xi_+=1$, $\xi_-=7$, $\xi_{\oplus}=7$, $\chi = -2$ k$_B$T, $\lambda_0=75$, and $\epsilon_r=5$. We can observe that the secondary satellite peak associated with the cation exchanges emerges at even larger negative voltages. This  is expected as a stronger ``bare" association constant increases the width of the voltage plateau. Additionally, the induced associations at modest positive potentials are further enhanced making the low potential differential capacitance peak even larger than the main case. After 0.076 V we would expect gel to be present in some parts of the EDL.}
     \label{fig:SI_cap_004}
\end{figure}

\begin{figure}[h!]
     \centering
     \includegraphics[width= .7\textwidth]{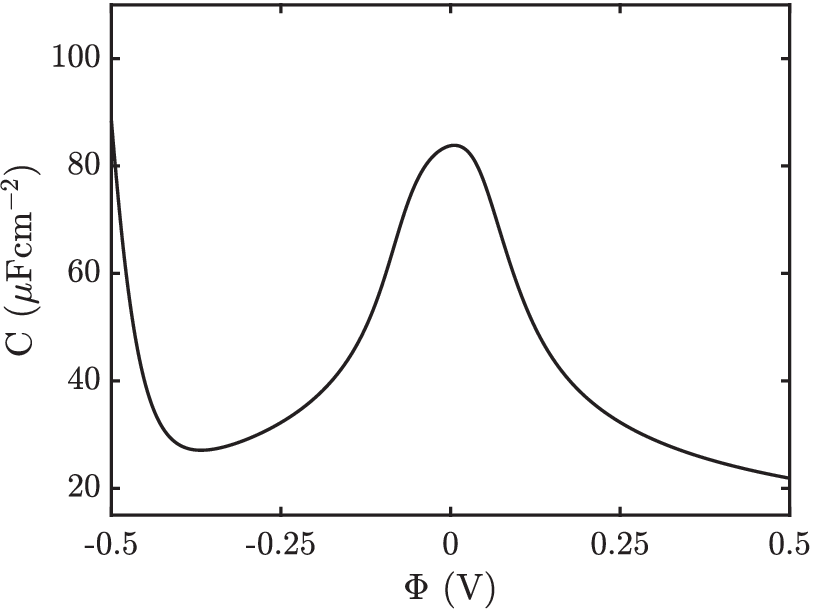}
     \caption{Differential capacitance of SiIL with a \underline{weaker regular solution interaction strength, $\chi$,} as a function of electrostatic potential. Here we use $x\textcolor{black}{_s} = 0.01$, $f_+=5$, $f_-=3$, $\xi_+=1$, $\xi_-=7$, $\xi_{\oplus}=7$, $\chi = -1$ k$_B$T, $\lambda_0=50$, and $\epsilon_r=5$. The secondary satellite peak associated with the cation exchanges emerges once again at an even higher voltage than the main case. This change is expected as a weaker regular solution interaction strength increases the width of the voltage plateau. Additionally we observe, as shown in Fig.~\ref{fig:SI_comp_005}, a decrease in the enhancement of aggregations at low positive potentials. This decrease appears to make the low potential differential capacitance peak significantly less prominent than the main case.}
     \label{fig:SI_cap_005}
\end{figure}

\begin{figure}[h!]
     \centering
     \includegraphics[width= 1\textwidth]{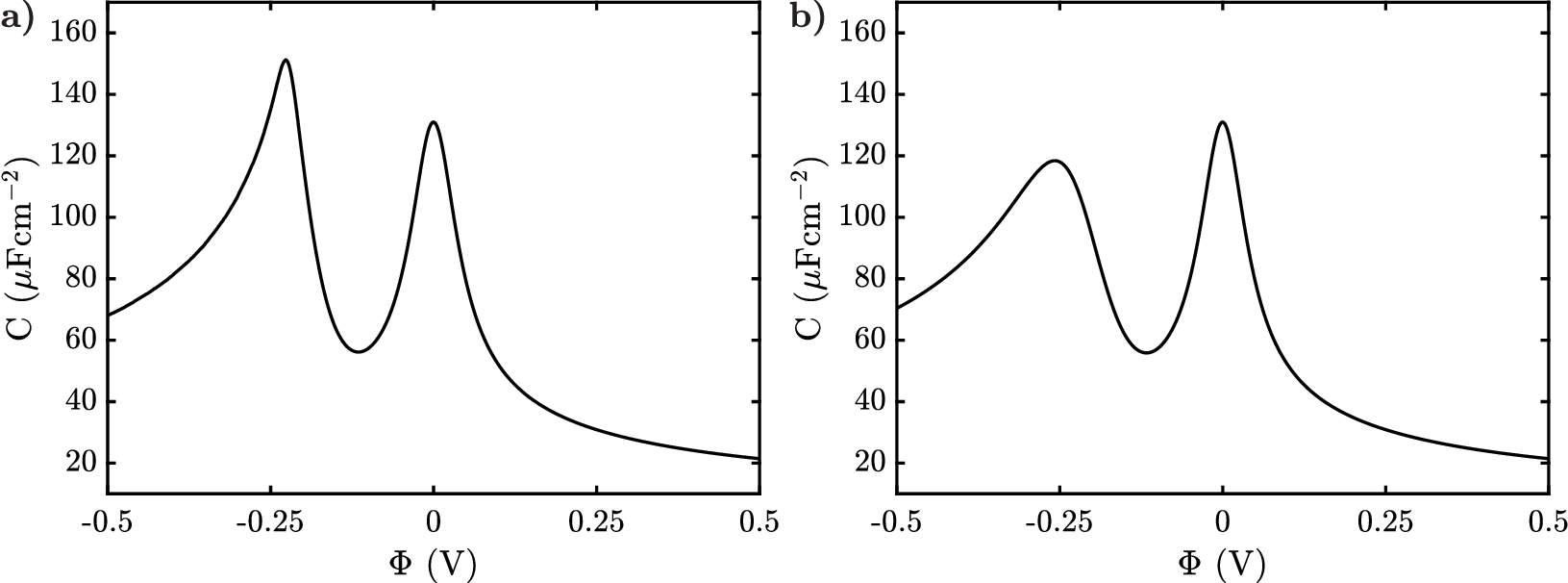}
     \caption{\textcolor{black}{Differential capacitance of SiIL without associations and regular solution interactions as a function of electrostatic potential. a) Incompressible model. b) Asymmetric 3-component Langmuir model~\cite{mceldrew2018}. Here we use $x_s = 0.01$, $\xi_+=1$, $\xi_-=7$, and $\xi_{\oplus}=7$ for both models. Here we can observe that the initial bell shape is much sharper in both cases compared to the main case. This influence may be due to the lack of associations and regular solution interactions compressing the initial bell shape. We can note that the secondary satellite peak associated with the cation exchange occurs in both models. }}
     \label{fig:SI_Filling_Cap}
\end{figure}

\begin{figure}[h!]
     \centering
     \includegraphics[width= 1\textwidth]{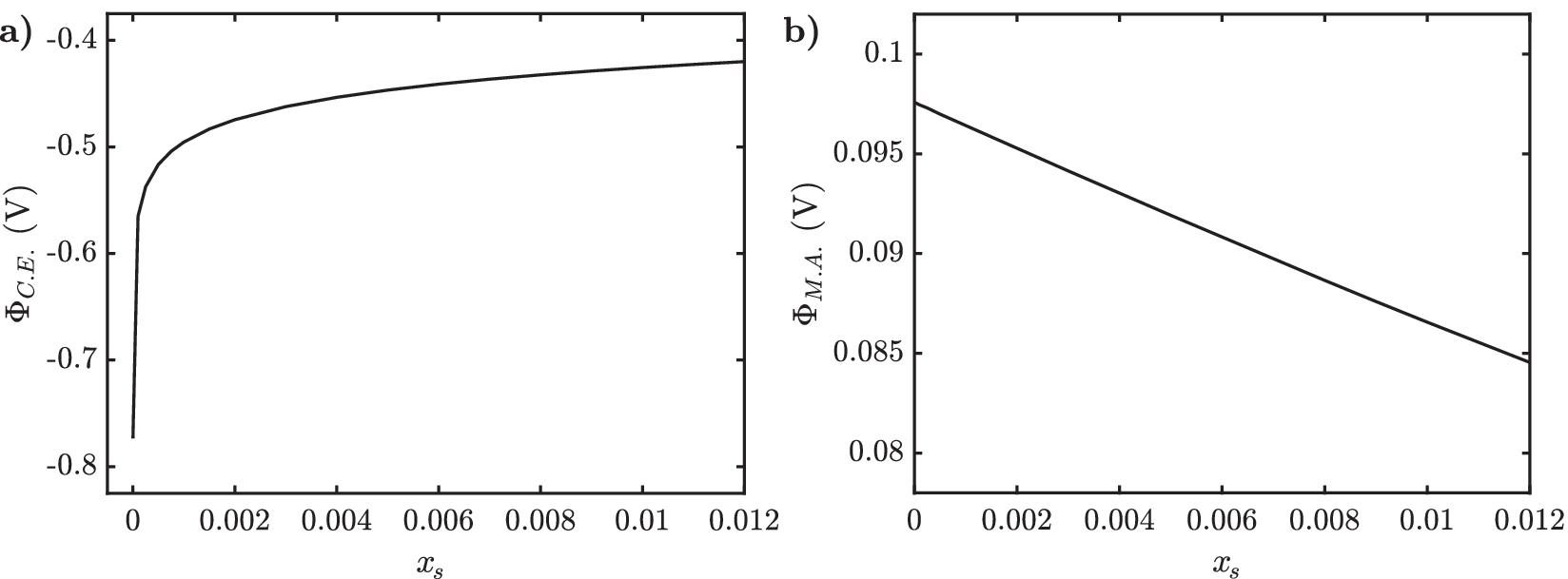}
     \caption{Voltage trends of SiIL as a function of the mole fraction of alkali metal salt. \textcolor{black}{a)} Cation exchange voltage\textcolor{black}{. b)} Maximum aggregation voltage. Here we use $f_+=5$, $f_-=3$, $\xi_+=1$, $\xi_-=7$, $\xi_{\oplus}=7$, $\chi = -2$ k$_B$T, and $\lambda_0=50$. \textcolor{black}{a)} Note as the mole faction of alkali metal salt goes to zero the cation exchange voltage diverges and at zero it does not exist. \textcolor{black}{b)} Note the maximum aggregation voltage also does not exist at zero.}
     \label{fig:SI_x_voltage}
\end{figure}

\begin{figure}[h!]
     \centering
     \includegraphics[width= 1\textwidth]{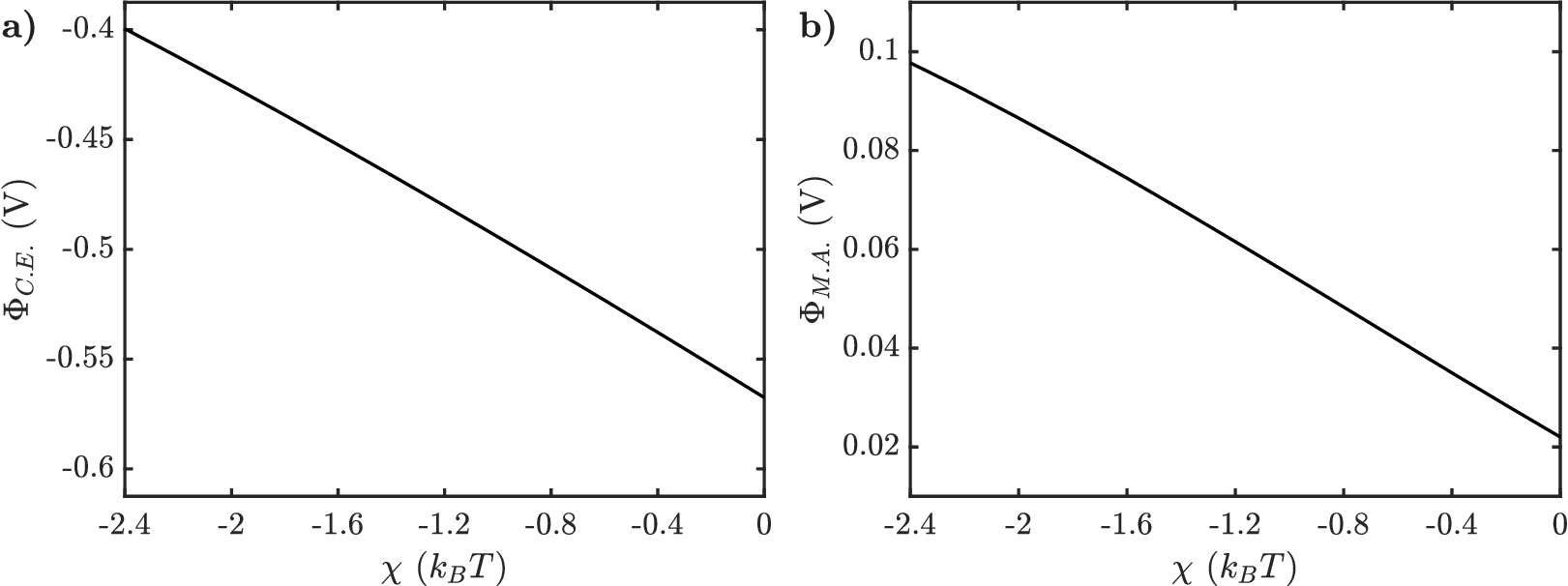}
     \caption{Voltage trends of SiIL as a function of the regular solution interaction strength, $\chi$. \textcolor{black}{a)} Cation exchange voltage\textcolor{black}{. b)} Maximum aggregation voltage. Here we use $x\textcolor{black}{_s} = 0.01$, $f_+=5$, $f_-=3$, $\xi_+=1$, $\xi_-=7$, $\xi_{\oplus}=7$, and $\lambda_0=50$.}
     \label{fig:SI_chi}
\end{figure}

\begin{figure}[h!]
     \centering
     \includegraphics[width= 1\textwidth]{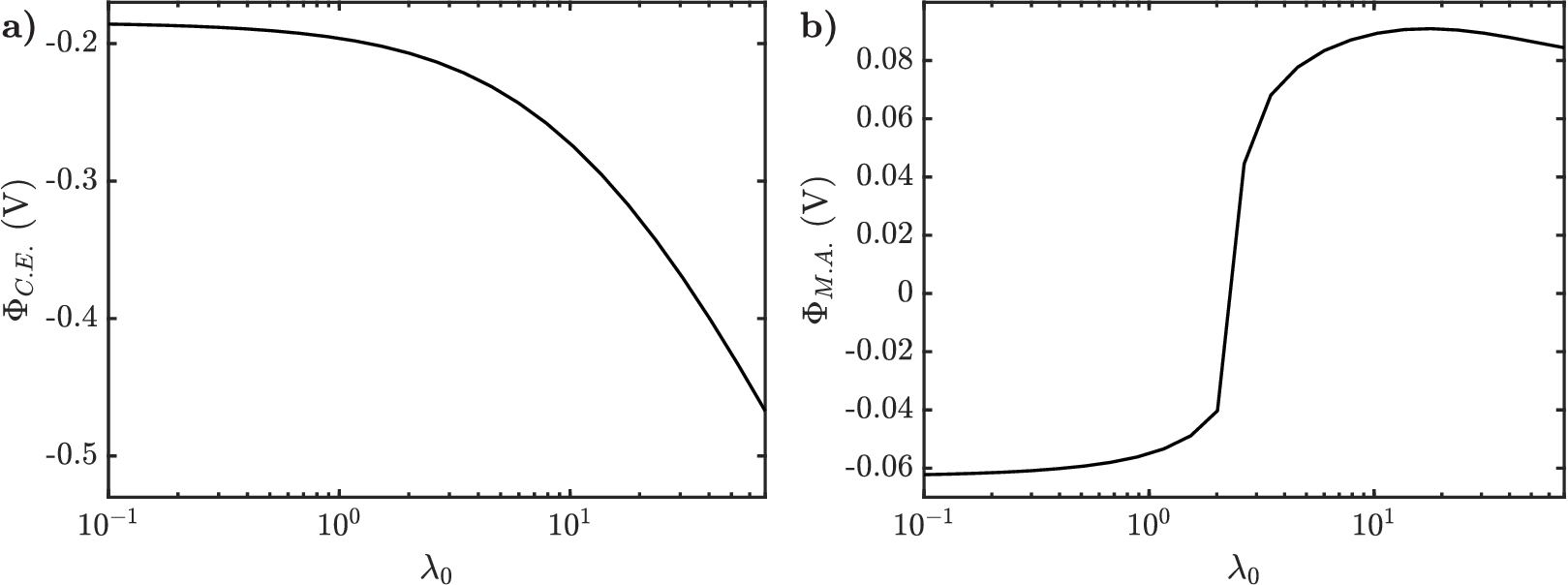}
     \caption{Voltage trends of SiIL as a function of the ``bare" association constant, $\lambda_0$. \textcolor{black}{a)} Cation exchange voltage\textcolor{black}{. b)} Maximum aggregation voltage. Here we use $x\textcolor{black}{_s} = 0.01$, $f_+=5$, $f_-=3$, $\xi_+=1$, $\xi_-=7$, $\xi_{\oplus}=7$, and $\chi = -2$ k$_B$T.}
     \label{fig:SI_lambda0}
\end{figure}

\clearpage
\section{Notes on system of equations and calculations}

In this section we highlight the system of equations used, how they can be implemented, and how to utilize them to reproduce the calculations shown in the main text and supplemental material.

\subsection{System of Equations}

As discussed in the main text we arrive at the following set of 6 equations:

\begin{gather*}
\bar{\phi}_{10} = \phi_{10}\exp(-e\beta\Phi + \Lambda) \\
\bar{\phi}_{01} = \phi_{01}\exp(e\beta\Phi- \beta\chi f_- (\bar{\phi}_{\oplus}-\phi_{\oplus}) + \xi_-\Lambda) \\
\bar{\phi}_{\oplus} = \phi_{\oplus}\exp\left(-e\beta\Phi + \beta\chi f_- \xi_{\oplus}\left\{c_{-}(1 - p_{-+})-\bar{c}_{-}(1 - \bar{p}_{-+})\right\} + \xi_{\oplus}\Lambda\right) \\
\bar{\phi}_+ + \bar{\phi}_- + \bar{\phi}_{\oplus} = 1 \\
\psi_{+}p_{+-}=\frac{1 + \lambda (\psi_- + \psi_{+}) - \sqrt{\left[1 + \lambda (\psi_- + \psi_{+})\right]^2-4 \lambda^2 \psi_- \psi_{+}} }{2 \lambda}\\
\psi_- p_{-+}=\frac{1 + \lambda (\psi_- + \psi_{+}) - \sqrt{\left[1 + \lambda (\psi_- + \psi_{+})\right]^2-4 \lambda^2 \psi_- \psi_{+}} }{2 \lambda}
\end{gather*}

\noindent We convert reduce this system of equations to only depend on ($\bar{\phi}_+$,$\bar{\phi}_-$,$\bar{\phi}_\oplus$,$\bar{p}_{+-}$,$\bar{p}_{-+}$,$\Lambda$) by asserting $\phi_{10} = \phi_{+}(1 - p_{+-})^{f_{+}}$ and  $\phi_{01} = \phi_-(1 - p_{-+})^{f_-}$ as mentioned in the main text. From this we obtain the following system of equations,

\begin{gather*}
\bar{\phi}_{+} = \frac{\phi_{10}}{(1 - \bar{p}_{+-})^{f_{+}}}\exp(-e\beta\Phi + \Lambda) \\
\bar{\phi}_{-} = \frac{\phi_{01}}{(1 - \bar{p}_{-+})^{f_{-}}}\exp(e\beta\Phi- \beta\chi f_- (\bar{\phi}_{\oplus}-\phi_{\oplus}) + \xi_-\Lambda) \\
\bar{\phi}_{\oplus} = \phi_{\oplus}\exp\left(-e\beta\Phi + \beta\chi f_- \xi_{\oplus}\left\{c_{-}(1 - p_{-+})-\bar{c}_{-}(1 - \bar{p}_{-+})\right\} + \xi_{\oplus}\Lambda\right) \\
\bar{\phi}_+ + \bar{\phi}_- + \bar{\phi}_{\oplus} = 1 \\
\psi_{+}p_{+-}=\frac{1 + \lambda (\psi_- + \psi_{+}) - \sqrt{\left[1 + \lambda (\psi_- + \psi_{+})\right]^2-4 \lambda^2 \psi_- \psi_{+}} }{2 \lambda} \\
\psi_- p_{-+}=\frac{1 + \lambda (\psi_- + \psi_{+}) - \sqrt{\left[1 + \lambda (\psi_- + \psi_{+})\right]^2-4 \lambda^2 \psi_- \psi_{+}} }{2 \lambda}
\end{gather*}

\noindent where the last three equations hold for both the EDL and bulk quantities. Other formulations and manipulations of this system of equations can be constructed and are equivalent to the one presented here so long as the alternative system of equations satisfy the original equations presented in the main text. With this and the bulk quantities calculated, as discussed in the main text, one can implement them into a numerical solver to obtain values for ($\bar{\phi}_+$,$\bar{\phi}_-$,$\bar{\phi}_\oplus$,$\bar{p}_{+-}$,$\bar{p}_{-+}$,$\Lambda$) at a given electrostatic potential ($\Phi$) and bulk properties ($\phi_+$,$\phi_-$,$\phi_\oplus$,$\lambda_0$,$\chi$).

\subsection{EDL Calculations}

To solve for profiles in the EDL as well as the differential capacitance, screening length, cation exchange voltage and maximum aggregation voltage, we use the following steps:

\begin{enumerate}
    \item First, we solve the our system of equations numerically for $\bar{\phi}_+$,$\bar{\phi}_-$,$\bar{\phi}_\oplus$,$\bar{p}_{+-}$,$\bar{p}_{-+}$,$\Lambda$ over a range of electrostatic potential values. This is done for a spectrum of dimensionless electrostatic potential ($\Phi/e\beta$) to allow one to create a sufficiently fine grid, for our purposes here a spacing of 0.001 was used. Here to simplify further steps, we stored the resulting fine maps for $\bar{\phi}_+$,$\bar{\phi}_-$,$\bar{\phi}_\oplus$,$\bar{p}_{+-}$,$\bar{p}_{-+}$,$\Lambda$, and dimensionless $\rho_e$, allowing for one to interpolated solutions to these quantities for a given electrostatic potential.
    \item Using the dimensionless $\rho_e$ map, we can then numerically solve the Poisson-Boltzmann equation to get a solution for the electrostatic potential profile in the EDL.
    \item The electrostatic potential profile in the EDL along with our interpolation maps allows us to predict profiles of the various quantities of interest in the EDL: dimensionless charge density, total volume fractions of each species, volume fraction of free cations, volume fraction free anions, volume fraction aggregates (could additional extract individual clusters volume fractions), association probabilities, and the product of association probabilities.
    \item In order to obtain our differential capacitance predictions, we first solved for the potential at the interface using a wide range of surface charge densities for our boundary conditions, for our work here we used a fine grid spacing of 0.0001 C/m$^2$. From this map, we constructed splines to calculate how the surface charge density depends on the potential at the interface. Using this we were able to calculate the differential capacitance numerically using finite differences.
    \item To obtain the screening length, we next applied a $\pm$ 0.0001 V electrostatic potential boundary condition at the surface. From the electrostatic potential profile in the EDL, we obtained the screening length by fitting the exponential decay for the profile, thus extracting the exponential decay constant for a range of mole fractions of alkali metal salt shown in Fig. 5\textcolor{black}{.a)}. In Fig. 5\textcolor{black}{.a)} it was constructed using a -0.0001 V electrostatic potential boundary condition. It is worth noting that the screening lengths obtained by $\pm$ 0.0001 V solutions are very similar. Additionally for the screening length plot in Fig. 5\textcolor{black}{.a)} and Fig.~\ref{fig:SI_screencap}\textcolor{black}{.a)} these profiles are within the pre-gel regime.
    \item The cation exchange voltage and maximum aggregation voltage utilized the method in step 1. For these plots, we utilize the values for the case displayed in the main text but individually varied the values of the mole fraction of alkali metal salt ($x\textcolor{black}{_s}$), regular solution interaction strength ($\chi$), and ``bare" association constant ($\lambda_0$) extracting the cation exchange voltage and maximum aggregation voltage for each case. For the cation exchange voltage plot shown here, we numerically solve for the larger negative voltage at which the spline fitted $\bar{\phi}_\oplus$ and $\bar{\phi}_+$ were equal. There can also be a second voltage at a smaller magnitude where they equal. For the voltage at the maximum aggregation, we extracted it directly by finding the minimum of the proximity of gelation.
\end{enumerate}

\bibliography{wise.bib}

\end{document}